\documentclass{aa} 
\usepackage[varg]{txfonts} 
\usepackage{natbib} 
\usepackage{graphicx}
\usepackage[bookmarks]{hyperref}
\hypersetup{pdfborderstyle={/S/U/W 1}}

\usepackage{multirow}
\usepackage{amsmath}

\usepackage{graphicx}
\usepackage{caption}
\usepackage{subcaption}

\bibpunct{(}{)}{;}{a}{}{,}

\renewcommand{\vec}{\boldsymbol} 
\newcommand{\matr}{\mathbf} 
 
\newcommand{\unit}{\mathrm} 
\newcommand{\name}{\mathrm} 
\newcommand{\dd}{\mathrm{d}} 

\newcommand{\appropto}{\mathrel{\vcenter{
			\offinterlineskip\halign{\hfil$##$\cr
				\propto\cr\noalign{\kern2pt}\sim\cr\noalign{\kern-2pt}}}}}

\begin{document}

\title{Formation of H$_2$-He substellar bodies in cold conditions:} 
\subtitle{Gravitational stability of binary mixtures in a phase transition} 

\author{A. F\"uglistaler \and D. Pfenniger} 
\institute{Geneva Observatory, University of Geneva, Sauverny, Switzerland\\ \email{andreas.fueglistaler@unige.ch}} 
\date{Received 16 July 2015/ Accepted  25 March 2016}

\abstract 
{ Molecular clouds typically consist of $3/4$ H$_2$, $1/4$ He and traces of heavier elements. In an earlier work we
    showed that at very low temperatures and high densities, H$_2$ can be in a phase transition leading to the formation of
    ice clumps as large as comets or even planets. However, He has very different chemical properties and no phase
    transition is expected before H$_2$ in dense interstellar medium (ISM) conditions. The gravitational stability of fluid
    mixtures has been studied before, but these studies did not include a phase transition.}
{ We study the gravitational stability of binary fluid mixtures with special emphasis on when one component is in a
    phase transition. The numerical results are aimed at applications in molecular cloud conditions, but the theoretical
    results are more general. }
{ First, we study the gravitational stability of van der Waals fluid mixtures using linearized analysis and examine
    virial equilibrium conditions using the Lennard-Jones intermolecular potential. Then, combining the Lennard-Jones and
    gravitational potentials, the non-linear dynamics of fluid mixtures are studied via computer simulations using the
    molecular dynamics code LAMMPS. }
{ Along with the classical, ideal-gas Jeans instability criterion, a fluid mixture is always gravitationally unstable if
    it is in a phase transition because compression does not increase pressure. However, the condensed phase fraction
    increases. In unstable situations the species can separate: in some conditions He precipitates faster than H$_2$, while
    in other conditions the converse occurs. Also, for an initial gas phase collapse the geometry is essential. Contrary to
    spherical or filamentary collapses, sheet-like collapses starting below 15\,K easily reach H$_2$ condensation conditions
    because then they are fastest and both the increase of heating and opacity are limited.}
{ Depending on density, temperature and mass, either rocky H$_2$ planetoids, or gaseous He planetoids form. H$_2$
    planetoids are favoured by high density, low temperature and low mass, while He planetoids need more mass and can form
    at temperature well above the critical value.}


\maketitle

\section{Introduction} 

Typically, the Milky Way molecular clouds consist of molecular hydrogen ($^1$H$_2$) and helium ($^4$He) in the
respective mass fraction of ${\sim}74\%$ and ${\sim}24\%$ and traces of heavier elements in the form of atoms,
molecules, and dust grains \citep{draine_physics_2011}. The He mass fraction is thus non-negligible.  Even though H$_2$
and He are by far the most abundant chemical components, they remain hardly detectable, and most of the time they are
inferred from CO emissions \citep{bolatto_co--h_2013}.  Thus, the dynamical and chemical processes associated with H$_2$
and He in molecular clouds are still poorly known, especially when considering sub-AU scales.

In \citet[][hereafter FP2015]{fuglistaler_substellar_2015}, we discussed substellar fragmentation including gravity in
single species fluids presenting a phase transition, such as very cold molecular hydrogen in molecular cloud conditions.
We showed that fluids in a phase transition (i.e. subject to a chemical instability) are anyway also gravitationally
unstable because any density fluctuation is not compensated by a pressure variation, but by a change in condensed matter
fraction. In phase transition conditions arbitrary small condensed clumps can form. The possibility of forming H$_2$ ice
clumps in the ISM, from grains, to comet-like bodies to rocky or gaseous planet-like bodies provides a scenario for
baryonic dark matter extending the scenario of \citet{pfenniger_is_1994-1,pfenniger_is_1994} towards micro-AU scales. 
However, since molecular clouds contain a substantial fraction of He, it is necessary to investigate how this component
might modify the findings of our previous study.

\begin{figure}[t] 
    \resizebox{\hsize}{!}{\includegraphics{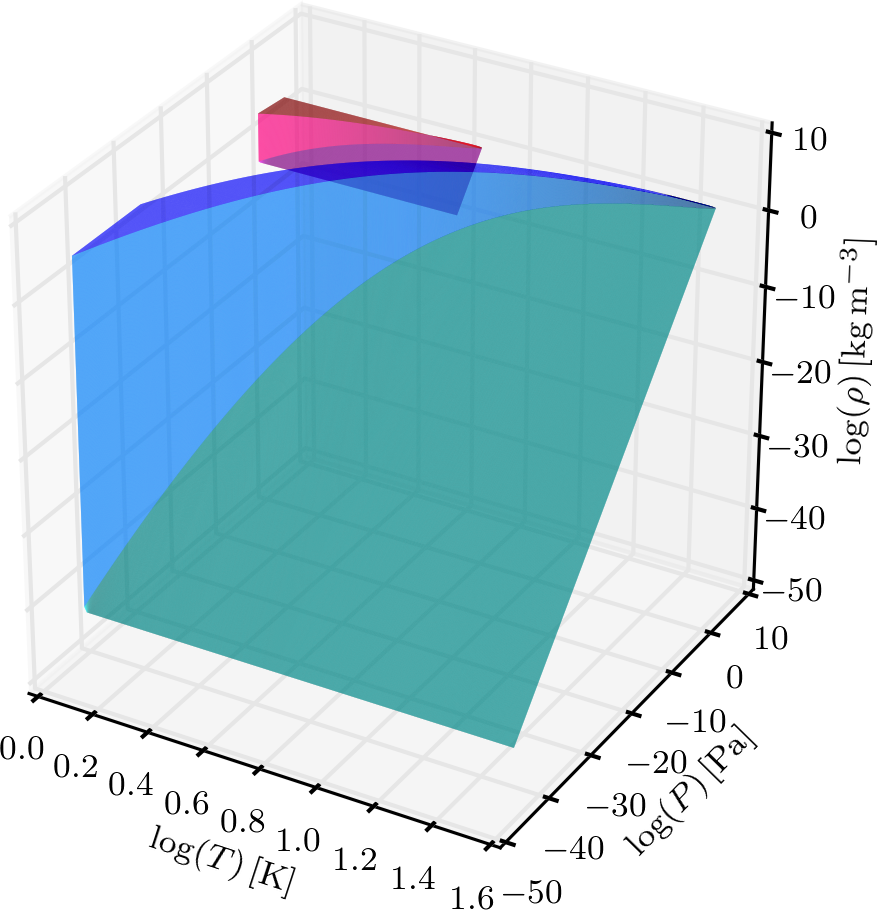}} 
    \caption{H$_2$ (blue) and He (red) phase diagrams.  For clarity, only a part of the upper, almost constant density
        condensed phases of both species and the low-density gas phase of He are shown. }
    \label{fig:cc} 
\end{figure}

Although at first sight from the chemical point of view both H$_2$ and He present an outer electronic shell made of two
electrons, their chemical properties differ markedly, mainly because of quantum physics. The individual properties of
H$_2$ and He are well known from laboratory data \citep{air_liquide_gas_1976} and shown in Fig.\ \ref{fig:cc}. H$_2$ and
He are in a phase transition when on the condensation wall linking the gaseous and solid or liquid phase.  He has a
lower critical temperature than H$_2$ ($5.2\,\unit{K}$ vs.\ $32.9\,\unit{K}$) and a lower critical pressure
($0.227\,\unit{MPa}$ vs.\ $1.286\,\unit{MPa}$). In the highly dynamical conditions present in molecular clouds, such as
supersonic turbulence \citep{elmegreen_interstellar_2004}, phase transition conditions may be reached thanks to a
combination of pressure increase and/or temperature decrease. In such a case, phase transition conditions are reached
for H$_2$ well before He. The conditions of phase transition of the mixture H$_2$-He may, however, change the
conclusions made in the single species case.

The properties of H$_2$-He mixtures has been mostly studied in detail for temperatures above the critical and at high
densities \citep[e.g.][]{streett_phase_1973,koci_study_2007,becker_ab_2014} and especially in conditions similar to gas
giant planets \citep{vorberger_hydrogen-helium_2007, saumon_equation_1995}. Taking quantum effects into account,
\citet{safa_equation_2008} calculate  the thermodynamic properties of H$_2$-He mixtures below critical temperature from
the known intermolecular potentials and obtain the critical point and stability of the mixture itself.

The gravitational stability of self-gravitating binary and multicomponent fluids has been studied by
\citet{grishchuk_gravitational_1981}, who showed that there can be only one unstable solution. If a fluid mixture is
gravitationally unstable then all components are affected. They note that in the case $\left(\partial P/\partial
\rho\right)_s=0$ (i.e. the sound-velocity formally vanishes), which is the case in a phase transition, the fluid is
always gravitationally unstable, but they do not go into more detail on that specific case. \citet{jog_galactic_1984,
    jog_two-fluid_1984} first discussed the stability of two-component disks. In a similar fashion
\citet{de_carvalho_onset_1995} studied oscillations and resonances in a binary fluid mixture.
\citet{volkov_gravitational_2000} discussed the stability of self-gravitating systems with a spectrum of particle masses
and consider rotating mediums.

When a phase transition occurs in the presence of external or internal gravity, the fluid dense phase may precipitate in
the form of rain, snow, or hail in the atmosphere, leading to a fragmentation that is impossible to describe with usual
hydrodynamic codes, in which a single phase in local thermal equilibrium is implicitly assumed. In FP2015 we showed that
method phase transition and precipitation can be simulated for a single species with molecular dynamics.  The possible
objects condensing from the gaseous phase can take various masses, typically covering the entire range from grains,
comets to planets or larger.  With two species with different molecular weights the number of precipitation scenarios
that can be envisioned increases.  Could it be, for example, that bodies form with a core made of solid H$_2$ surrounded
with an atmosphere of H$_2$ and He or that a solid H$_2$ crust floats on a gaseous He core?

To answer such questions we use the same molecular dynamics code as in FP2015 just adding a second species, and scaling
the particles properties to the respective properties of H$_2$ and He. We control the finite number resolution effects
by performing simulations over a range from $1.25\cdot 10^5$ to $80\cdot 10^5$ particles.  We restrict the
investigations to the simplest set-up combining gravity with molecular dynamics.  To control gravitational instability,
we investigate a single plane-parallel collapse in one direction of a periodic cube, where the initial temperature and
density are simulation parameters.  As explained in Sect.\ \ref{subsec:ppc} and Appendix B, the collapse geometry
(sheet-, filament-, or point-like) is crucial to reach phase transition conditions starting from typical ISM conditions.
Sheet-like collapses (pancakes) can indeed lead temporarily to very dense conditions without much heating, contrary to
the other cases.

\section{Gravitational stability of a fluid mixture}

In a fluid consisting of $K$ different components $i$, the total mixture number density is $n = \sum_i n_i$, the mixture
mass density is $\rho = \sum_i \rho_i$ with $\rho_i = m_i n_i$, and the mixture pressure is $P = \sum_i P_i$. In the
case of an ideal gas, Dalton's law states $P = \sum_i x_i P_i$. Each component has a molecular fraction $x_i = n_i / n$
and a mass fraction $w_i = m_i/m$ with $m = \sum_i x_i m_i$.

The notion of global temperature in a system with long-range forces is an unsettled topic as the key assumption of
extensivity in thermodynamics breaks down in long-range force systems \citep[e.g.][]{padmanabhan_statistical_1990}. When
dealing with particle systems we can however always define the temperature as proportional to the residual kinetic
energy when the bulk translational, expansional, and rotational velocities are subtracted, be it globally or locally.
Strictly, this definition is operational and useful only if the velocity distribution is unimodal and its second order
moment exists.  Further detailed discussion about this topic would be out of scope, as in this article we consider
either global or local temperatures for particle systems with no or negligible amount of ordered motion, so the stated
temperature is equivalent to the particle kinetic energy. The high degree of collisionality in molecular interactions
ensures the rapid destruction of any initial correlations leading to the convergence towards thermal states.

\subsection{Jeans instability}

Considering a fluid mixture as a one-component fluid using average quantities such as the density $\rho$ and pressure
$P$, the Jeans instability criterion (see App. \ref{app:oc}) would be
\begin{equation}
\label{equ:kJi}
k^2 < k_\name{J}^2 \equiv {4\pi G \rho \over (\partial P / \partial \rho)_s}  \ ,
\end{equation}
where $k$ is the wavenumber, $k_\name{J}$ the critical Jeans wavenumber, and $G$ the gravitational constant. This is,
however, inappropriate if each component has a different mean square velocity, which is the case for an isothermal fluid
with different molecular masses.

In order to correctly predict the stability of a many-component fluid mixture with different densities $\rho_i$ and
partial pressure $P_i$, each component has to be treated individually: the Grishchuk-Zeldovich criterion (see App.
\ref{app:tc}), which is the sum of each component Jeans' criteria, reads
\begin{equation}
\label{equ:kM}
k^2 < k_\name{GZ}^2 \equiv \sum_i {4\pi G\rho_i\over {\partial P_i / \partial \rho_i}} \ .
\end{equation}
\subsubsection{Ideal gas mixture}
A fluid far from the condensed phase can be approximated with the ideal gas law
\begin{eqnarray}
P &=& n k_\name{B}T \ , \\
\left(\partial P \over \partial \rho\right)_s &=& \gamma {k_\name{B}T \over m} \ ,
\end{eqnarray}
where $k_\name{B}$ is the Boltzmann constant, $\gamma$ the adiabatic index, and $m$ the molecular mass. The partial
pressures can be calculated using Dalton's law. Equ.\ (\ref{equ:kM}) becomes
\begin{equation}
\label{equ:kGZid}
k_\name{GZ,id}^2 = {4\pi G \over \gamma k_\name{B} T}\sum_i {n_i m_i^2} \ ,
\end{equation}
where id stands for ideal gas. This equation differs from Equ.\ (\ref{equ:kJi}), which in the ideal gas case becomes
\begin{equation}
k_\name{J,id}^2 = {4\pi G \over \gamma k_\name{B}T}n m^2 \ .
\end{equation}
In the case where all components have the same temperature $k_\name{GZ,id} \geq k_\name{J,id}$ independent of the
molecular fractions $x_i=n_i/n$ and molecular mass $m_i$. In the case of a H$_2$-He mixture, the maximum
$k_\name{GZ,id}^2/k_\name{J,id}^2 = 1.12$ at $x_\name{H_2} = 0.67$.

\subsubsection{van der Waals fluid mixture}
\label{subsec:vdW}

When approaching a phase transition, the ideal gas law does not take into account condensations and does not yield
correct values anymore. We showed in FP2015 that the van der Waals equation of state \citep{van_der_waals_remarks_1910}
describes a phase transition rather well provided that the Maxwell construct is taken into account
\citep{clerk-maxwell_dynamical_1875,johnston_thermodynamic_2014}, i.e.
\begin{eqnarray}
P_\name{r} &=& \frac{8T_\name{r}}{\frac{3}{n_\name{r}}-1} - 3n_\name{r}^2  \ , \label{equ:vdw}\\
\left({\partial P_r \over \partial \rho}\right)_s  &=& 
\gamma\left({24 T_r \over m (n_r-3)^2} - {6n_r\over m}\right) \ ,
\end{eqnarray}
in gaseous and solid/liquid form with the reduced values $P_\name{r}  = {P/ P_\name{c}}$, $T_\name{r} = {T /
    T_\name{c}}$, $n_\name{r} = {n/ n_\name{c}}$ and the critical pressure, temperature and density $P_\name{c}$,
$T_\name{c}$ and $n_\name{c}$. In the case of a phase transition, $P_\name{r} = \name{const}$ (Maxwell construct) and
$({\partial P_r / \partial \rho})_s = 0$.

The Maxwell line is very similar to the laboratory condensation line for H$_2$ in a $T-P$ diagram as can be seen in
Fig.\ \ref{fig:maxlab}, but is rather off for He, especially at low temperatures. As in the astrophysical context,
correctly representing the H$_2$ phase transition is essential for our study.  A H$_2$ phase transition always occurs at
a lower pressure-temperature ratio than for He.

In the phase transition regime $(\partial P / \partial \rho)_s = 0$, varying density allows the pressure to remain
constant. This is a crucial property for this work, since gravitational contraction is no longer compensated by pressure
increase. We show in App.\ \ref{app:tc} that a two-component fluid is always gravitationally unstable as soon as
$(\partial P_i / \partial \rho_i)_s = 0$ for any component $i$. In a similar fashion, the same can be deduced for
$n$-component fluids \citep{grishchuk_gravitational_1981}.

\subsection{Plane-parallel collapse}
\label{subsec:ppc}

\begin{figure}[t] 
    \resizebox{\hsize}{!}{\includegraphics{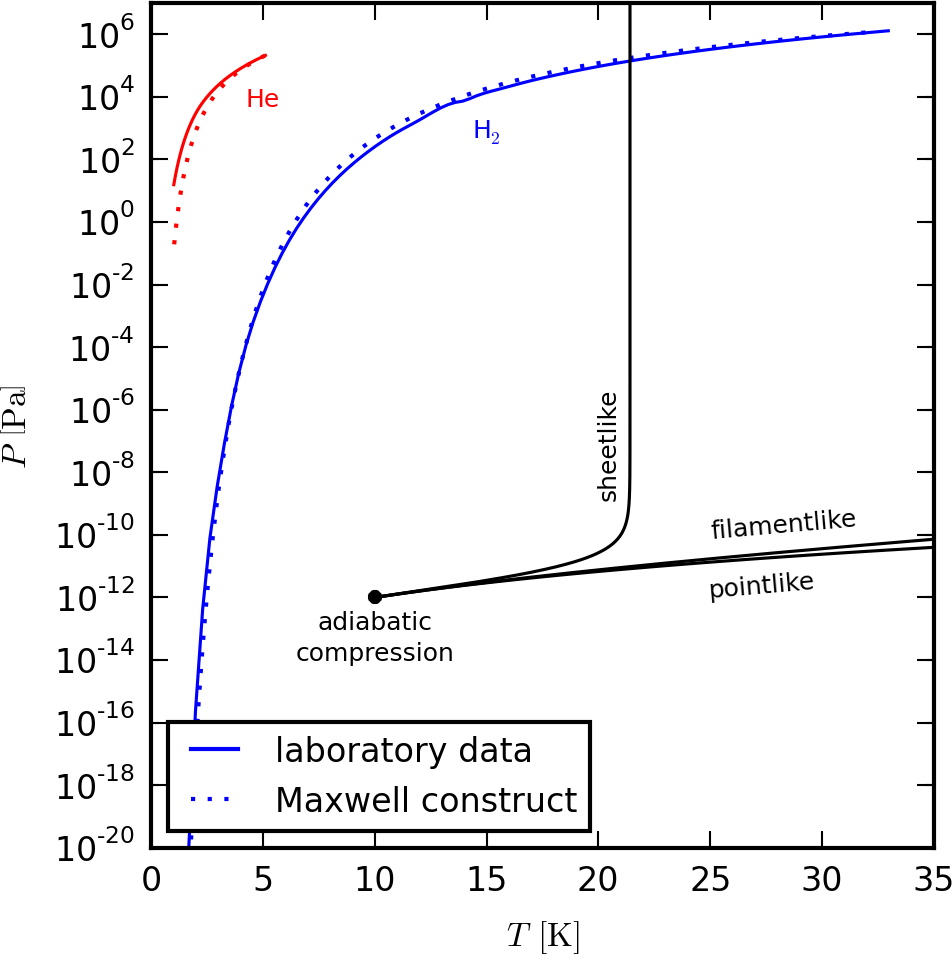}}
    \caption{H$_2$ and He laboratory data and van der Waals vapour curve derived from Maxwell construct. Adiabatic
        compression curves of an initial sphere to sheet-, filament- and point-like geometries for interstellar conditions
        ($T = 10\,\unit{K}$, $P = 10^{-12}\,\unit{Pa}$) are shown, as explained in Appendix B.  Sheet-like collapses are
        allowed to reach the phase transition regime even without cooling. Cooling would displaces the curves to lower
        temperature. }
    \label{fig:maxlab}
\end{figure}

\citet{lin_gravitational_1965} and \citet{zeldovich_gravitational_1970} show that a plane-parallel collapse, leading to
a sheet-like geometry, is a faster collapse than filament- or point-like geometries. This has been confirmed using
numerical simulations by \citet{shandarin_shape_1995}.

In addition, as shown in App. \ref{app:ge}, the adiabatic matter compression of a sheet-like geometry leads to a finite
increase of potential energy, leading to a maximum relative temperature increase of only 2.1, while the energy diverges
logarithmically in a filament-like geometry  and as $Z^{1/3}$ in a point-like geometry, where
$Z=\rho_\name{final}/\rho_\name{initial}$ is the density compression.  This can be seen in Fig.\ \ref{fig:maxlab} which
shows how a sphere  moves in this diagram when adiabatically compressed towards a sheet, filament, or point initially in
interstellar conditions ($T = 10\,\unit{K}$, $P = 10^{-12}\,\unit{Pa}$). Whereas the temperature is quickly increasing
with filament- and point-like geometries, in the sheet-like geometry it only rises to ${\sim} 21\,\unit{K}$, which is
well below the 33\,K critical temperature of H$_2$.

When taking  radiative cooling into account, the temperature increase by contraction is even smaller.  In App.\
\ref{app:c} we show that the opacity of a sheet-like collapse is barely increasing. If the initial medium is
transparent, the final sheet is also transparent.  On the other hand, in filament- and point-like collapses the opacity
increases approximately as a power $1/2$ or $2/3$ of compression, quickly reaching  a full opacity regime able to stop
the collapse.

\subsection{Lennard-Jones mixtures}
\label{subsec:LJ}

The Jeans instability of Equ.\ (\ref{equ:kM}) requires thermal equilibrium and is simplified by only considering linear
perturbations. It does not predict its non-linear evolution when unstable.  This is a motive to use molecular dynamical
simulations with a Lennard-Jones potential in addition to gravity for studying such non-linear phenomena.

The Lennard-Jones  potential
\begin{equation}
\Phi_{\name{LJ}}(r) = 4\frac{\epsilon}{m}\left[\left(\sigma \over r\right)^{12}  
- \left(\sigma \over r\right)^{6}\right] 
\end{equation}
reproduces the van der Waals equation of state when in equilibrium conditions using the following relations
\citep{caillol_critical-point_1998}:
\begin{eqnarray}
\label{equ:Tc}
T_\name{c} &=& 1.326 \, {\epsilon \over k_\name{B}} \ , \\
\label{equ:nc}
n_\name{c} &=& 0.316 \, \sigma^{-3} \ , \\
\label{equ:Pc}
P_\name{c}  &=& 0.157\, { \epsilon \over k_\name{B}\sigma^{3} } \ .
\end{eqnarray}
For a fluid mixture, we use the usual Lorentz-Berthelot combining rule for the Lennard-Jones potential between two
molecules
\begin{equation}
\Phi_{\name{LJ}}(r_{ij}) = 4\frac{\epsilon_{ij}}{m_i}\left[\left(\sigma_{ij} \over r_{ij}\right)^{12}  
- \left(\sigma_{ij} \over r_{ij}\right)^{6}\right] \label{equ:phiij} \ ,
\end{equation}
where $\sigma_{ij} = {1 \over 2}(\sigma_i + \sigma_j)$ and $\epsilon_{ij} = \sqrt{\epsilon_i \epsilon_j}$
\citep{lorentz_ueber_1881,berthelot_sur_1898}. The most accurate mixing rules for energy and distance are
\citep{banaszak_mixing_1995,chen_comparison_2001}
\begin{eqnarray}
\label{equ:epsx}
\sigma^3 &=& \sum_i^K \sum_i^K x_i x_j \sigma_{ij}^3\ , \\
\label{equ:sigx}
\epsilon &=& {\sum_i^K \sum_i^K x_i x_j \epsilon\sigma_{ij}^3 \over \sigma^3}  \ .
\end{eqnarray}
Since $T_\name{c} \propto \epsilon$ and $n_\name{c} \propto \sigma^{-3}$, we have $n_\name{c} / n_\name{c, \alpha} =
(\sigma / \sigma_\name{\alpha})^{-3}$ and $T_\name{c} / T_\name{c, \alpha} = \epsilon / \epsilon_\name{\alpha}$. Using
Equ.\ (\ref{equ:vdw}), we find $P_\name{c} = (8/3) T_\name{c}n_\name{c}$, and therefore ${P_{c} / P_{c,\alpha}} =(\sigma
/ \sigma_\name{\alpha})^{-3} \cdot ({\epsilon / \epsilon_\name{\alpha}})$.

As in the ideal gas case, using these mixing properties to calculate $k_\name{J}$ in analogy to Equ.\ ({\ref{equ:kJi})
instead of using Equ.\ (\ref{equ:kM}) would lead to different results. In the ideal gas case, the biggest difference is
${\sim}10\%$, but in the present case of a van der Waals fluid the difference can be much bigger, for example
$k_\name{J}$ always returns a finite number if the critical temperature of the mixture $T_\name{c} > 1$, whereas
$k_\name{GZ}$ can nevertheless be infinite if one of the components is in a phase transition.

\subsubsection{Binary mixture}

\begin{figure}[t] 
  \resizebox{\hsize}{!}{\includegraphics{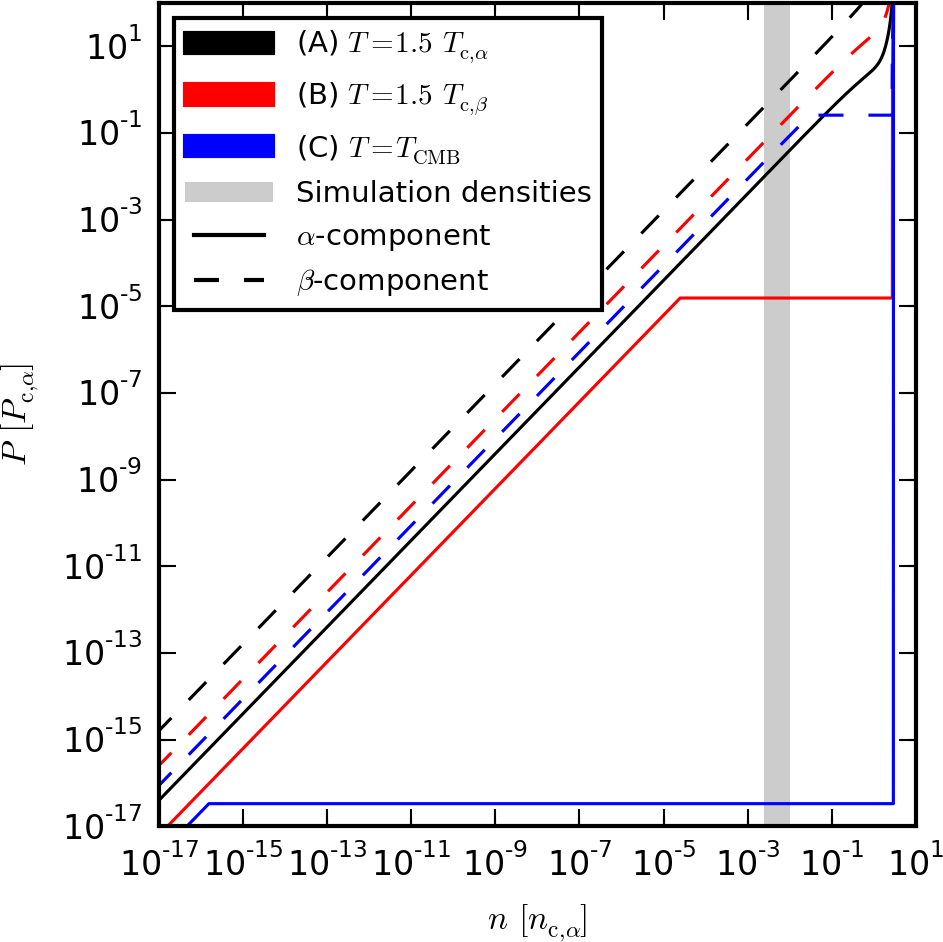}}
  \caption[]{van der Waals phase diagram with Maxwell construct for the H$_2$ -- He mixture, each species considered as
      independent.}
  \label{fig:cases}
\end{figure}

Considering a fluid consisting of two components $\alpha$ and $\beta$, we define
\begin{eqnarray}
\theta &=& {T_\name{c,\beta} \over T_\name{c, \alpha}} = {\epsilon_\beta \over \epsilon_\alpha} \ , \label{equ:theta} \\
\nu &=& {n_\name{c,\beta} \over n_\name{c, \alpha}} = \left({\sigma_\beta \over \sigma_\alpha}\right)^{-3} \ , \\
\mu &=& {m_\beta \over m_\alpha} \ . \label{equ:mu}
\end{eqnarray}
The following Lennard-Jones properties can be derived using Equ.\ (\ref{equ:phiij} -- \ref{equ:mu}):
\begin{eqnarray}
\sigma_{\alpha\beta} &=& {1\over 2}\,\sigma_\alpha\left(1 + \nu^{-1/3}\right) \ ,\\
\epsilon_{\alpha\beta} &=& \epsilon_\alpha\theta^{1/2} \ ,
\end{eqnarray}
and the following van der Waals mixture properties:
\begin{eqnarray}
  {n_\name{c,x} \over n_\name{c,\alpha}} &=& \left[x_\alpha^2 
+ {x_\alpha x_\beta\over 4}\left(1 + \nu^{-1/3} \right)^3 
+ {x_\beta^2 \over \nu}\right]^{-1} \ ,\\
{T_\name{c, x}\over T_\name{c, \alpha}} &=& {n_\name{c,x} \over n_\name{c,\alpha}}\left[x_\alpha^2 \,\nu^{1/3}
+ {x_\alpha x_\beta\,\theta^{1/2} \over 4}\left(1 + \nu^{-1/3} \right)^3 + 
{x_\beta^2\theta \over \nu} \right] \ ,\\
  \label{equ:px}
  {P_\name{c,x} \over P_\name{c, \alpha}} &=& {n_\name{c,x} \over n_\name{c,\alpha}}\cdot {T_\name{c, x}\over T_\name{c, \alpha}} \ ,
\end{eqnarray}
where $x_\beta = 1-x_\alpha$. Without loss of generality we choose $\theta < 1$. 
\medskip

Three different cases can be distinguished, as shown in Fig.\ \ref{fig:cases}:
\begin{enumerate}[(A)]
\item $ T_\name{c, \beta} < T_\name{c, \alpha} <T$ \\
  Having the temperature above both critical values, neither component can be in a phase transition and $k_\name{GZ}$ is
  always finite.
\item $T_\name{c, \beta} < T < T_\name{c, \alpha}$ \\
  $\beta$ is still above the critical temperature and cannot be in a phase transition, but $\alpha$ may or may not be in
  a phase transition depending on $n_\alpha$.
\item $T < T_\name{c, \beta} < T_\name{c, \alpha}$ \\
  Having the temperature below both critical temperatures, both components can be in a phase transition. At equal
  component number density, $\alpha$ is faster in a phase transition, but theoretically, at low $n_\beta$, $\beta$ could
  be in a phase transition without $\alpha$, but this hardly ever happens in reality.
\end{enumerate}
These three cases are studied using computer simulations (see Table \ref{tab:aT}).

\subsubsection{Hydrogen-helium mixture}

The critical temperature of H$_2$ and He are $32.97$ and $5.19\unit{K}$, whereas their usual Lennard-Jones $\epsilon$
values are $36.4$ and $10.57\,k_\name{B}\unit{K}$. This is in conflict with the temperature conversion of Equ.\
(\ref{equ:theta}), as $\theta_\name{H_2-He} = 6.35$ using critical temperatures whereas $\theta_\name{H_2-He} = 3.44$
using the Lennard-Jones $\epsilon$ values. The same is the case to a lesser degree for the critical density.

All performed simulations are molecule independent, but $\theta$, $\nu$, and $\mu$ need to be defined. These values were
set using $T_\name{c}$, $n_\name{c}$, and the molecular mass of laboratory He and H$_2$ data
\citep{air_liquide_gas_1976}. The goal of this article is to understand the role of a secondary component in a fluid
presenting a phase transition together with gravity. In molecular clouds, the most likely case of a phase transition is
$T_\name{c, H_2} >  T > T_\name{c, He}$, thus it is important to have a correct $(T_\name{c,He} / T_\name{c,H_2})$
fraction.

\subsubsection{Virial theorem}
\label{subsubsect:vir}

\begin{figure}[t] 
	\resizebox{\hsize}{!}{\includegraphics{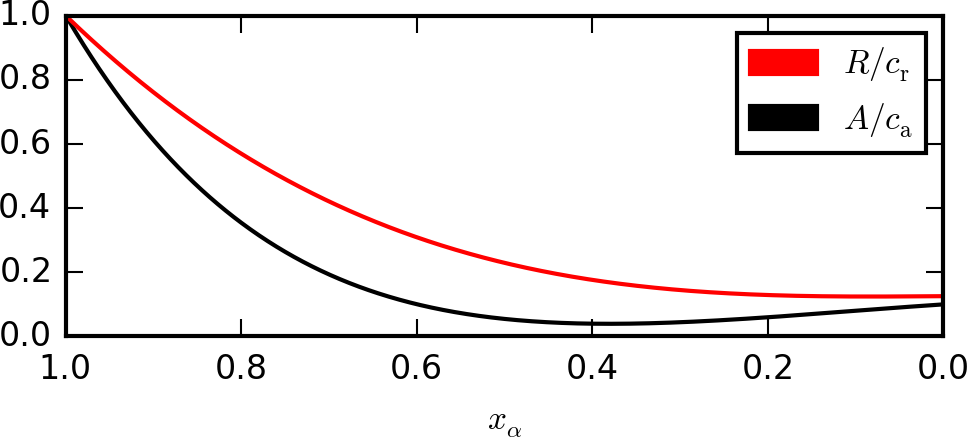}}
	\caption{$R/c_\name{r}$ and $A/c_\name{a}$values as a function of $x_{\alpha}$ for a H$_2$-He mixture.}
	\label{fig:RA}
\end{figure}

In FP2015, the Lennard-Jones potential has been decomposed in attractive and repulsive terms: $\Phi_\name{LJ} =
\Phi_\name{a} + \Phi_\name{r}$. In a binary mixture, we have to further distinguish the one-component terms
$\Phi_{\alpha^2}$ and $\Phi_{\beta^2}$, and the cross-terms $\Phi_{\alpha\beta}$ and $\Phi_{\beta\alpha}$ (see Equ.\
\ref{equ:phiij}).

In analogy with Equ.\ (5) of FP2015, the virial theorem becomes as follows:
\begin{equation} 
0 = \underbrace{2 E_\name{kin} 	+ 12E_\name{r}}_{>0} 
+ \underbrace{6E_\name{a}	+ E_\name{G} }_{<0} \ .
\end{equation}
There are two negative, attractive terms, and two positive, repulsive terms. If the attractive and repulsive terms
equalize each other, the system is in virial equilibrium.

In the case of a homogeneous density and species distribution of mass $M$, the energy terms are
\begin{eqnarray}
E_\name{kin} &=&   \phantom{-}{3 k_\name{B}T\over 2m} M \ ,\\
E_\name{r} &=& \phantom{-}R{\epsilon_\alpha \sigma_\alpha^{12}n^4\over m}M \ , \\
E_\name{a} &=& - A{\epsilon_\alpha \sigma_\alpha^{6}  n^2\over m}M \ , \\
E_\name{G} &=&  -G\, f_\name{G}\left(n, m\right)M^{5/2}  \ , 
\end{eqnarray}
with $f_\name{G} > 0$ depending on the geometry. The repulsive and attractive constants are
\begin{eqnarray}
R &=&  c_\name{r} \left[ x^5_\alpha  + \theta\,\nu^{-4}x^5_\beta  
+ {\theta^{1/2} \over 2^{12}}\left(1 + \nu^{-1/3}\right)^{12}\left(x_\alpha x_\beta^4 + x_\alpha^4 x_\beta\right)\right] \ , \\
A &=&  c_\name{a} \left[x^3_\alpha   + \theta\,\nu^{-2}x^3_\beta   
+ {\theta^{1/2}\over 2^{6}}\left(1 + \nu^{-1/3}\right)^{6}~\left(x_\alpha x_\beta^2 + x_\alpha^2 x_\beta\right)\right]  \ ,
\end{eqnarray}
where the lattice coefficients $c_\name{r} $ and $c_\name{a}$ depend on the specific crystal lattice (FP2015), which are
of importance for the solid phase.  Figure \ref{fig:RA} shows $R$ and $A$ as functions of an abundance number fraction.

The above terms are for a fluid with no spatial separation of the species, i.e. a fluid in its initial state. The terms
predict how an unstable fluid evolves. However, once a phase transition or a gravitational collapse happens, the species
may separate. In that case, the terms have to be calculated for each species independently, using $x_\alpha = 1$ and
$0$, respectively.

\subsubsection{Unvirializable densities}

\begin{figure}[t] 
        \resizebox{\hsize}{!}{\includegraphics{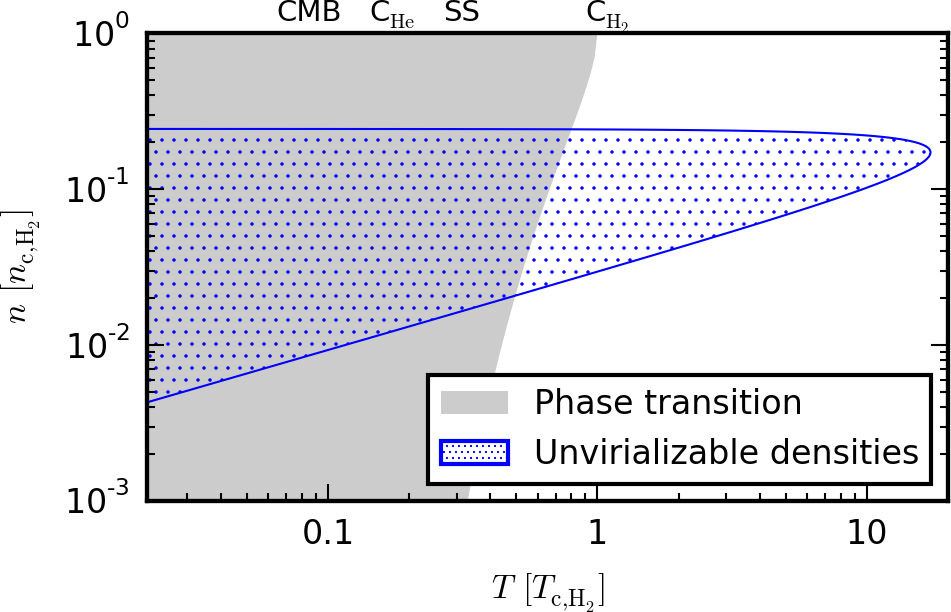}}
        \caption{van der Waals phase transition and unvirializable density for a binary mixture with $x_\alpha=0.75$.}
        \label{fig:virvdw}
\end{figure}

The sum of the Lennard-Jones and kinetic terms must be positive as the gravitational term is negative: $E_\name{G} < 0 <
E_\name{r} + E_\name{a} + E_\name{kin}$. This is the case if
\begin{equation}
2 A \epsilon_\alpha\left({\sigma_\alpha^3 n}\right)^2 \leq k_\name{B}T 
+4 R\epsilon_\alpha\left({\sigma_\alpha^3n}\right)^4\ . \label{equ:AR}
\end{equation}
There can be no virial equilibrium for any mass $M>0$, if the attractive Lennard-Jones term dominates the kinetic and
repulsive terms. We define the unvirializable density domain as follows:
\begin{eqnarray}
\mathcal{D} &\equiv&  \{n : n_{-} < n < n_{+}\} \nonumber \\
n_{\pm}^2 &=& \sigma^{-6}{1 \pm \sqrt{1-4\,{R \over A^2}{k_\name{B}T \over \epsilon}}\over 4\,{R\over A}} \ .
\end{eqnarray}
If the term in the square root is negative, there is no real solution and therefore no unvirializable densities. This is
the case if
\begin{eqnarray}
T > T_\name{max} \equiv {A^2 \over 4 R}{\epsilon \over k_\name{B}} \ . 
\end{eqnarray}
Figure \ref{fig:virvdw} shows the $n_\pm$ values and the domain of the van der Waals phase transition for H$_2$-He
mixtures with a molecular fraction of $x_\alpha = 0.75$. There are unvirializable densities above the critical
temperature $T_\name{c}$ up to $T_\name{max}$ even though no phase transition is possible.

The domain of phase transition does not change a lot for the different $x_\name{H_2}>0$, as the H$_2$ phase transition
temperature remains the same and the density changes as $n_x=x\cdot n$. It is at much lower temperatures for
$x_\name{H_2}=0$, as there is no H$_2$ anymore and the phase transition domain switches to He.

The evolution of fluids with unvirializable densities depends whether they are in a phase transition or not. If these
fluids are in a phase transition, there is a gravitational instability independent of $M$ (see \ref{subsec:vdW}), which
leads to a collapse and the formation of bodies of small mass. If they are not in a phase transition, there is only a
gravitational collapse above a certain mass $M$ (see Equ.\ \ref{equ:kM}). Below that mass, clumps form, which leads to
an augmentation of the kinetic and repulsive Lennard-Jones terms until Equ.\ (\ref{equ:AR}) is fulfilled, at which point
an equilibrium is reached and the fluid may remain stable.

\subsubsection{Dynamical friction}

When discussing gravitational collapses, the concept of dynamical friction \citep{chandrasekhar_dynamical_1943} is
important, since heavy objects may condensate from the gas and start to precipitate, i.e. move with respect to the gas
in the local gravity field.  Considering a uniform density and Maxwellian velocity distribution, the dynamical friction
of a heavy object of mass $m_\name{h}$ reads
\begin{equation}
{\dd \vec{v}_\name{h} \over \dd t} = 
-{4\pi G^2  m_\name{h}\,\rho \log(\Lambda)  \over v_\name{h}^3}
\left[\name{erf}(X)-{2X \over \sqrt{\pi}}\exp\left(-X^2\right)\right]\vec{v}_\name{h} \ ,
\end{equation}
where $\log(\Lambda)$ is the Coulomb logarithm and $X = v_\name{h}/\sqrt{2\sigma_v}$. For the qualitative analysis
needed in this work, it is enough to state
\begin{equation}
{\dd \vec{v}_\name{h} \over \dd t} \propto -m_\name{h} \ .
\end{equation}
This can be used in different cases. First, He is twice as heavy as H$_2$ and can therefore be considered a heavy object
and  in some conditions lead to sediment faster than H$_2$.  Secondly, when H$_2$ is in a phase transition, H$_2$ ice
grains may sediment faster than He.

\section{Method} 

For all of the simulations, the Large-Scale Atomic/Molecular Massively Parallel Simulator (LAMMPS) is used
\citep{plimpton_fast_1995}. The use of its short-range Lennard-Jones solver and long-range gravitational solver,
super-molecules, and the rRESPA time integrator are discussed in FP2015.

We recall the following super-molecule properties, where $\eta$ is the number of molecules per super-molecule,
\begin{eqnarray}
m_{\mathrm{SM}} &=& \eta\, m \ , \\
\epsilon_\mathrm{SM} &=& \eta \,\epsilon \ ,\\ 
\sigma_\mathrm{SM} &=& \eta^{1/3}\sigma \ , 
\end{eqnarray}
where $m$, $\sigma$, and $\epsilon$ are the values for one molecule. The gravitational constant, described in molecular
dynamics units ($\sigma = \epsilon = m = 1$), is
\begin{equation}
\label{equ:Geta}
G_\name{SM} = {G m^2 \over \sigma \epsilon}\eta^{2/3} \ .
\end{equation}
In order for the gravitational force between two super-molecules to be consistently small compared to the intermolecular
forces, $\eta$ should satisfy the following constraint:
\begin{equation} 
\label{equ:xglj}
\eta^\frac{2}{3} \ll {24\epsilon\sigma \over G\,m^2} 
\left( r_\name{c}^{-5} -  2r_\name{c}^{-11} \right) \ ,
\end{equation}
where $r_\name{c}$ is the cut-off radius in $\sigma$ units (set to $r_\name{c}=4$ in the simulations). 

Two molecules are considered as bound in LAMMPS if their distance is smaller than $1.3625 \sigma$. A clump of bound
molecules can be either gaseous or condensed, dependent whether its temperature is above or below the critical value.

Initially, the fluid is uniformly distributed in a periodic cubic box.  To reproduce the most generic plane-parallel
collapse first, as explained in Sect.\ \ref{subsec:ppc}, a velocity perturbation in form of a small plane sinusoidal
wave in the $x$ direction is superposed to the fluid's Maxwellian velocity distribution. The perturbation strength is of
$1\%$; see FP2015 for how the perturbation is calculated.

\subsection{Units}

All simulations are performed in dimensionless units and only the ratios of physical quantities matter.  The initial
properties of a fluid are the ratios of its temperature to the critical value $T_\name{c}$, its number density to the
critical value $n_\name{c}$, the Lennard-Jones constant ratios $\theta$, $\nu$, the mass ratio $\mu$, the molecular
fraction $x_\alpha$ and the gravitational potential strength $\gamma_\name{G}$. The needed molecule parameters were set
in accordance to laboratory H$_2$ and He values
\begin{eqnarray}
\theta &=& {T_\name{c, He} \over T_\name{c,H_2}} = 0.157 \ ,\\
\nu &=& {n_\name{c, He} \over n_\name{c,H_2}} = 1.12 \ ,\\
\mu &=& {m_\name{He} \over m_\name{H_2}} = 2 \ .
\end{eqnarray} 
The time unit is defined as the particle crossing time for the box of length $L$, i.e.
\begin{equation} 
\tau = \frac{L}{V}  \ ,
\end{equation}
where $V^2 = 3N k_\name{B} T/\sum_i m_i$, $i=1 \ldots N$.

The gravitational constant strength is measured by a factor $\gamma_\name{J}$ relative to the ideal gas Jeans limit
strength $G_\name{J}$, defined as
\begin{eqnarray}
\label{equ:GJ}
 \gamma_\name{J} &=& {G \over G_\name{J}} \ , \nonumber \\
 G_\name{J} &=& {\pi\gamma k_\name{B}T \over  L^2\sum\limits_{j=\alpha,\beta}{n_j m_j^2}}  \ .
\end{eqnarray}

\subsection{Visualization}

\begin{figure}[t] 
        \resizebox{\hsize}{!}{\includegraphics{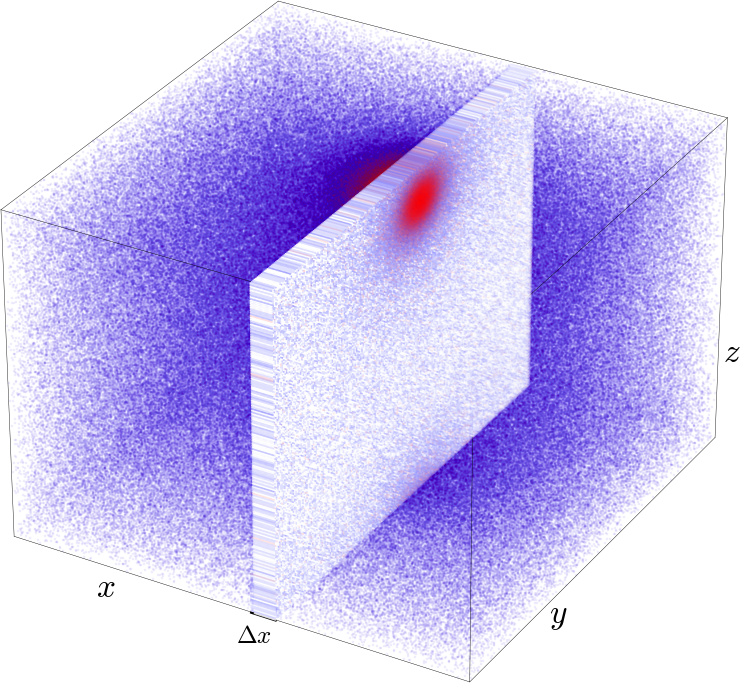}}
        \caption{Two-dimensional density map of three-dimensional space.}
        \label{fig:3dPlot}
        \end{figure}

\begin{figure}[t] 
	\centering
	\resizebox{.75\hsize}{!}{\includegraphics{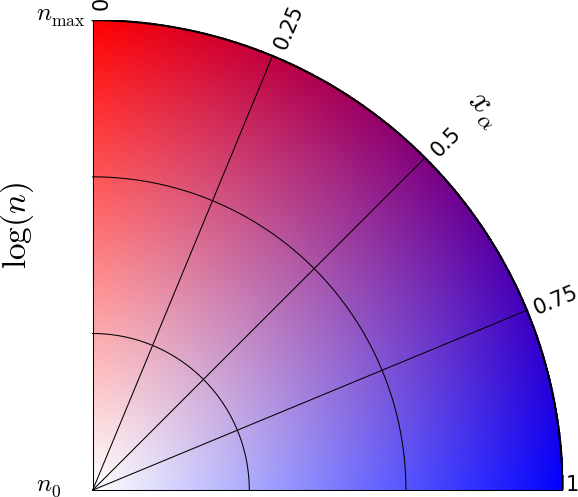}}
	\caption{Colour mapping of density map.}
	\label{fig:cmap}
\end{figure}
 
In order to visualize the particle snapshots, two-dimensional number density maps are used. The introduced perturbation
is along the $x$-axis, thus the sheet-like collapse is parallel to the $yz$-plane. For that reason, the density map
shows the $y$- and $z$-axes where most of the relevant events can be observed. When showing all $N$ particles, smaller
aggregates are washed out and barely visible, which is why only a slice of the whole simulation box is shown. Figure
\ref{fig:3dPlot} shows how this slice is selected: The highest number density is determined in the $x$ direction in
order to be centred around the collapse region. From there, the slice width $\Delta x$ is calculated in order to contain
$N_\name{slice} = f_\name{slice}N$ particles. In that way, the slice width $\Delta x$ differs for every snapshot, but
always contains the same number $N_\name{slice}$ of particles.

The number density $n$ and number fraction $x_\alpha$ are represented by brightness and colour. As at low density the
brightness is maximum and the colour is simply white, the colour map is best visualized in polar coordinates, where $n$
is the radius and $x_\alpha$ the angle. Figure \ref{fig:cmap} shows the colour map used, for better contrasts, the
brightness is represented in logarithmic scale.

\begin{table*}[ht] 
	\caption{Simulation parameters} 
	\label{tab:aT} 
	\centering 
	\begin{tabular}{l| l | l l |l l |l | l | l l | l l | l l} 
		\hline\hline 
		Name & $x_\alpha$ & \multicolumn{2}{c|}{$n$} & \multicolumn{2}{c|}{$T$} & $\gamma_\name{J}$ & $N_\name{tot}$ & 
		\multicolumn{2}{c|}{$M_\name{tot}/M_\oplus$\tablefootmark{a}\tablefootmark{b}} &
		\multicolumn{2}{c|}{$L/R_\oplus$\tablefootmark{a}\tablefootmark{b}} &
		\multicolumn{2}{c}{$\tau/\unit{kyrs}$\tablefootmark{a}}
		\rule{0pt}{2.6ex}\rule[-1.2ex]{0pt}{0pt}\\
		&&$[n_\name{c,\alpha}]$ & $[\unit{m^{-3}}]$\tablefootmark{a} & $[T_\name{c,\alpha}]$ & 
		$[\unit{K}]$\tablefootmark{a} &&&  \tablefootmark{(f)} & \tablefootmark{(g)} &
		\tablefootmark{(f)} & \tablefootmark{(g)} & \tablefootmark{(f)} & \tablefootmark{(g)}\\
		\hline 
		A10& $1.0$ & \multirow{5}{*}{$10^{-2}$} & \multirow{5}{*}{$9.3\cdot 10^{25}$} & \multirow{5}{*}{$1.5$} & 
		\multirow{5}{*}{$49.5$} & \multirow{5}{*}{$0$, $0.5$, $1.5$}  & \multirow{5}{*}{$100^3$} & 
		$0.34$& $1.7$& $27$ & $47$ & $0.86$ & $1.5$ \rule{0pt}{2.6ex}\\ 
		A75 & $0.75$ &&&&&&& $0.094$ & $0.49$ & $16$ & $28$  & $0.58$& $1.0$ \\ 
		A50 & $0.5$  &&&&&&& $0.039$ & $0.2$ & $11$ & $20$ & $0.45$ & $0.78$  \\ 
		A25 & $0.25$ &&&&&&& $0.019$ & $0.1$ & $8.6$ & $15$ & $0.36$ & $0.64$ \\ 
		A00 & $0.0$  &&&&&&& $0.011$ & $0.056$ & $6.8$& $12$  & $0.31$ & $0.53$ \\ 
		\hline
		A75$_\name{S}$ &\multirow{2}{*}{$0.75$} & $10^{-2}$ & $9.3\cdot 10^{25}$ & 
		\multirow{2}{*}{$1.5$} & \multirow{2}{*}{$49.5$} & $1.5$ & $50^3$ -- $160^3$ & 
		\multicolumn{2}{c|}{$0.49$} & \multicolumn{2}{c|}{$18$}& \multicolumn{2}{c}{$1.0$}\rule{0pt}{2.6ex}\\ 
		A75$_{01}$ && $10^{-1}$ &  $9.3\cdot 10^{26}$ &&& $0$, $1.5$ & $100^3$ & 
		\multicolumn{2}{c|}{$0.16$} & \multicolumn{2}{c|}{$9$} & \multicolumn{2}{c}{$0.29$}\\ 
		\hline
		B10 & $1.0$ & \multirow{5}{*}{$10^{-2}$} & \multirow{5}{*}{$9.3\cdot 10^{25}$} & \multirow{5}{*}{$0.24$} & 
		\multirow{5}{*}{$7.8$\tablefootmark{c}} &  \multirow{5}{*}{$0$, $0.5$, $1.5$}  & 
		\multirow{5}{*}{$80^3$} & $0.021$ & $0.11$ & $11$& $18$  & $0.86$ & $1.5$ \rule{0pt}{2.6ex}\\
		B75& $0.75$ &&&&&&& $0.0059$ & $0.031$ & $6.5$ & $11$ & $0.58$ & $1.0$\\
		B50& $0.5$  &&&&&&& $0.0024$ & $0.013$ & $4.5$ & $7.9$ & $0.45$ & $0.78$ \\ 
		B25& $0.25$ &&&&&&& $0.0011$ & $0.0063$ & $3.4$ & $5.9$ & $0.36$ & $0.63$\\ 
		B00& $0.0$  &&&&&&& $0.00068$ & $0.0035$ & $2.7$ & $4.7$ & $0.31$ & $0.53$\\ 
		\hline
		B75$_{\gamma}$& \multirow{2}{*}{$0.75$} & $10^{-2}$ & $9.3\cdot 10^{25}$ & \multirow{2}{*}{$0.24$} & 
		\multirow{2}{*}{$7.8$} & $0.5$ -- $1.5$ & \multirow{2}{*}{$80^3$} & 
		\multicolumn{2}{c|}{$0.0059$ -- $0.031$} & \multicolumn{2}{c|}{$6.5$ -- $11$} & 
		\multicolumn{2}{c}{$0.58$ -- $1.0$}\rule{0pt}{2.6ex}\\
		B75$_{01}$ && $10^{-1}$ & $9.3\cdot 10^{26}$ &&& $0$, $1.5$ &&
		\multicolumn{2}{c|}{$0.0097$} & \multicolumn{2}{c|}{$3.6$} & \multicolumn{2}{c}{$0.32$}\\
		\hline
		C10 & $1.0$ & \multirow{5}{*}{$10^{-2}$} & \multirow{5}{*}{$9.3\cdot 10^{25}$} & \multirow{5}{*}{$0.082$} & 
		\multirow{5}{*}{$2.7$\tablefootmark{d}}& \multirow{5}{*}{$0$, $0.5$, $1.5$}  & 
		\multirow{5}{*}{$80^3$}& $0.0043$ & $0.022$ & $6.3$ & $11$ & $1.5$ & $0.86$  \rule{0pt}{2.6ex}\\
		C75& $0.75$ &&&&&&& $0.0012$ & $0.0063$ & $3.8$ & $6.6$ &  $1.0$ & $0.58$   \\
		C50& $0.5$  &&&&&&& $0.00049$ & $0.0026$ & $2.7$ & $4.6$ &  $0.78$ & $0.45$  \\ 
		C25& $0.25$ &&&&&&& $0.00025$ & $0.0013$ & $2.0$ & $3.5$ &  $0.63$ & $0.36$  \\ 
		C00& $0.0$  &&&&&&& $0.00014$ & $0.00072$ & $1.6$ & $2.8$ &  $0.53$ & $0.31$  \\ 
		\hline		
		SSM$_{01}$ & \multirow{3}{*}{$0.84$} & $10^{-1}$ & $9.3\cdot 10^{26}$ &  \multirow{3}{*}{$0.3$} & 
		\multirow{3}{*}{$10$}& $1.05$ & \multirow{3}{*}{$80^3$} &
		\multicolumn{2}{c|}{$ 0.012$\tablefootmark{e}} & \multicolumn{2}{c|}{$3.9$} &
		\multicolumn{2}{c}{$0.3$}\rule{0pt}{2.6ex}\\
		SSM$_{02}$ && $10^{-2}$ & $9.3\cdot 10^{25}$ &&& $0.48$ &&\multicolumn{2}{c|}{$ 0.012$\tablefootmark{e}} &
		\multicolumn{2}{c|}{$8.5$}& \multicolumn{2}{c}{$0.65$}\\
		SSE$_{04}$ && $10^{-4}$ & $9.3\cdot 10^{23}$ &&& $1.98$ &&\multicolumn{2}{c|}{$1.0$} &
		\multicolumn{2}{c|}{$171$}& \multicolumn{2}{c}{$13$}\\
		\hline
	\end{tabular} 
	\tablefoot{All simulations are initially perturbed by a small plane sinusoidal wave in the $x$-direction as in FP2015.\\
		\tablefoottext{a}{Considering $\alpha = $ H$_2$ and $\beta = $ He.}
		\tablefoottext{b}{$\oplus\equiv$ Earth}
		\tablefoottext{c}{$T = 1.5T_\name{c, He}$.}
		\tablefoottext{d}{$T = T_\name{CMB}$.}
		\tablefoottext{e}{$M = M_\name{Moon}$.}
		\tablefoottext{f}{$\gamma_\name{J} = 0.5$.}
		\tablefoottext{g}{$\gamma_\name{J} = 1.5$.}
	}
\end{table*}

\section{Simulations}

In FP2015, we simulated fluids with only one component. We introduced the terms {comets} for clumps that are principally
bound by the Lennard-Jones potential and {planetoids} for clumps that are principally bound by gravity. If a fluid is in
a phase transition, it is important to distinguish between a strong gravitational potential above the ideal gas Jeans
criterion and a weak gravitational potential below it. In the strong gravity case, a gravitational collapse happens,
leading to the formation of a hot, gaseous planetoid. Phase transitions only happen at the beginning as the fluid heats
up above the critical temperature where no solid comets can form.

In the weak gravity case, no gravitational collapse happens and solid comets form thanks to the phase transition. The
comets attract each other gravitationally, which leads to the formation of a solid planetoid. During the comet
aggregation, the number of bound molecules does not rise. This means that the planetoid only captures comets and no
single molecules. Therefore, the fraction of bound molecules is identical to the number of molecules that underwent
phase transition.

In this article, we compare fluids with different molecular fractions by keeping the physical properties alike (constant
$T/T_\name{c,\alpha}$ and $n/n_\name{c,\alpha}$). By decreasing $x_\alpha$ the mean mass per molecule $m$ increases,
since $m_\alpha < m_\beta$. It is therefore not possible to have the same number of molecules per super-molecule $\eta$,
the same super-molecule mass $M_\name{SM}= m\eta$, and the same gravity $G_\name{J}\propto m^{2}\eta^{2/3}$ at the same
time.

Since we want to study the reaction of fluids with different $x_\alpha$ above and below the ideal gas Jeans criterion,
we need to ensure that $\gamma_\name{J}$ is $>1$ or $<1$ in the compared simulations. For that reason, neither the total
mass nor the number of molecules remains the same when comparing fluids with different $x_\alpha$, while
$\gamma_\name{J}$ remains the same.  Since $G \appropto m^{-2}$ (Equ.\ \ref{equ:GJ}) and $\eta^{2/3}\propto m^{-2}G$
(Equ.\ \ref{equ:Geta}), then $\eta\appropto m^{-6}$ and thus $M = m\eta \appropto m^{-5}$. Therefore, changing $m$ by a
factor $2$ leads to a total mass factor decrease of $32$.

Different cases are studied: $T = 1.5 T_\name{c, \alpha}$, $T = 1.5 T_\name{c,\beta}$ and $T =
(T_\name{CMB}/T_\name{c,H_2})\,T_\name{c, \alpha}$ with $n = 10^{-2} n_\name{c,\alpha}$ and $x_\alpha = 1$, $0.75$,
$0.5$, $0.25$, and $0$. These initial parameters are shown in Fig.\ \ref{fig:cases}. In addition, different densities
are used for simulations B75 and C75 to compare cases with $n > n_-$ with $n < n_-$ (see Sect.\ \ref{subsubsect:vir}).
We also study solar system abundances for different total masses and number densities. The simulations are summarized in
Table \ref{tab:aT}.

The simulations with $\gamma_\name{J} < 1$ were run until a steady-state solution was reached. Most simulations with
$\gamma_\name{J} > 1$ were stopped once a planetoid formed, which typically happens after ${\sim}5\tau$. Even if the
resulting fluid did not reached a steady state then, no further developments are expected, as checked in FP2015. A few
simulations were run for $15\tau$ to confirm this. At $t=5\tau$, the planetoid of B75 with $\gamma = 1.5$ is $15\%$
hotter than average, whereas unbound molecules are $5\%$ colder. After another $10\tau$, however, at $t=15\tau$, the
temperature differences are below $1\%$. Similarly, at $t=5\tau$, there are very strong regional temperature differences
of $\geq 10\%$ considering subdomains of $1\%$ volume. At $t=15\tau$, they are $\leq 5\%$.

\subsection{Above critical temperatures}

In this Section, we consider fluids where both $T_\alpha$ and $T_\beta$ are above the critical temperature. The number
densities corresponding to the different abundance number fractions can be seen in Fig.\ \ref{fig:cases}.

\subsubsection{Time evolution}

\begin{figure}[ht] 
        \resizebox{\hsize}{!}{\includegraphics{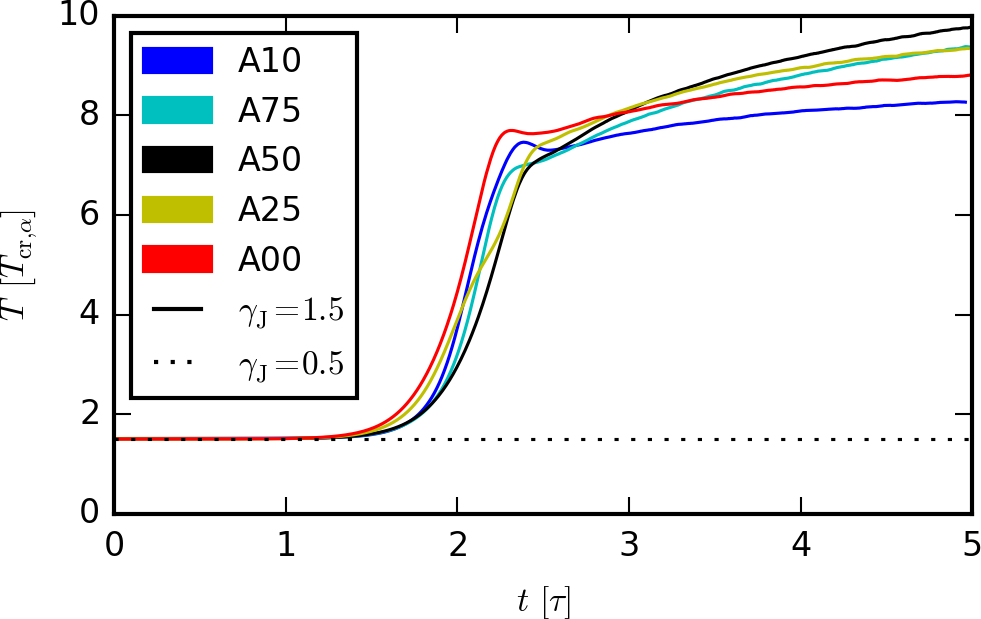}}
        \caption[]{Global temperature as a function of time of the A simulations with $\gamma_\name{J} = 1.5$ and $0.5$.}
        \label{fig:A}
\end{figure}

\begin{figure*}[t] 
    {\includegraphics[width=17cm]{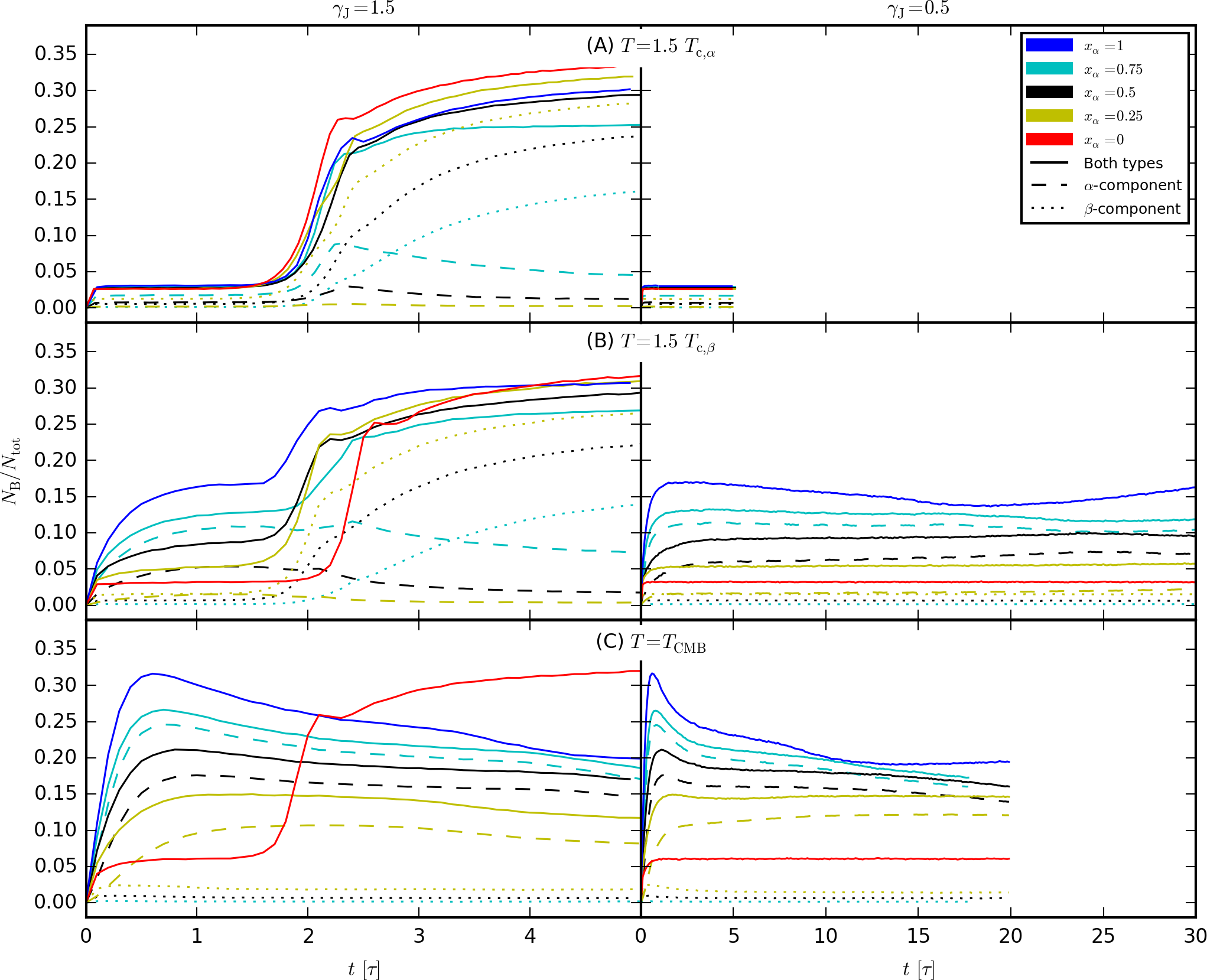}}
    \caption[]{Evolution of the number of bound molecules as a function of time of the A, B, and C simulations with
        $\gamma_\name{J}=1.5$ and $0.5$. The simulations with $\gamma_\name{J}=0.5$ are stopped after $5\tau$, $30\tau$, and
        $20\tau$ for the A, B, and C simulations, respectively, when no further significant development is expected.}
    \label{fig:simNB}
\end{figure*}

Figure \ref{fig:A} shows the global temperature evolution of the A simulations (see Table \ref{tab:aT}). In all
simulations, the weakly self-gravitating fluids with $\gamma_\name{J} = 0.5$ and the fluids without gravity are very
similar and do not react significantly to the velocity perturbation. On the other hand, as expected for the sufficiently
self-gravitating fluids with $\gamma_\name{J} = 1.5$ the introduced perturbation rises exponentially. The reaction time
is similar for all number fractions, but the mixtures reach slightly higher temperatures. Keeping in mind that all
simulations have the same $\gamma_\name{J}$ value, but the total mass differs with $M_1/M_2 = (m_1/m_2)^{-5}$.

To get a deeper understanding of the internal processes, we need to distinguish between the $\alpha$ and $\beta$
component. Figure \ref{fig:simNB}A shows the fraction of bound molecules in the simulations as a function of time. No
phase transition happens above the critical temperature, therefore no comets form in the simulations with
$\gamma_\name{J} = 0$ and $0.5$. With $\gamma_\name{J} = 1.5$, the gravitational collapse leads to the formation of a
planetoid. In its centre, the gravitational pull is strong enough to keep the molecules bound even though the
temperature is well above the critical value.

The $\beta$-molecules, as they are twice as heavy as the $\alpha$-molecules, fall faster into the forming planetoid.
Indeed, even in A75, with only $x_\beta = 0.25$, the fraction of bound $\beta$-molecules surpasses the fraction of bound
$\alpha$-molecules for $\tau > 2$.

\subsubsection{Planetoid formation}
\begin{figure*}[p] 
        \centering      
        \begin{subfigure}[b]{0.49\textwidth}
        \centering
        {\includegraphics[width=\textwidth]{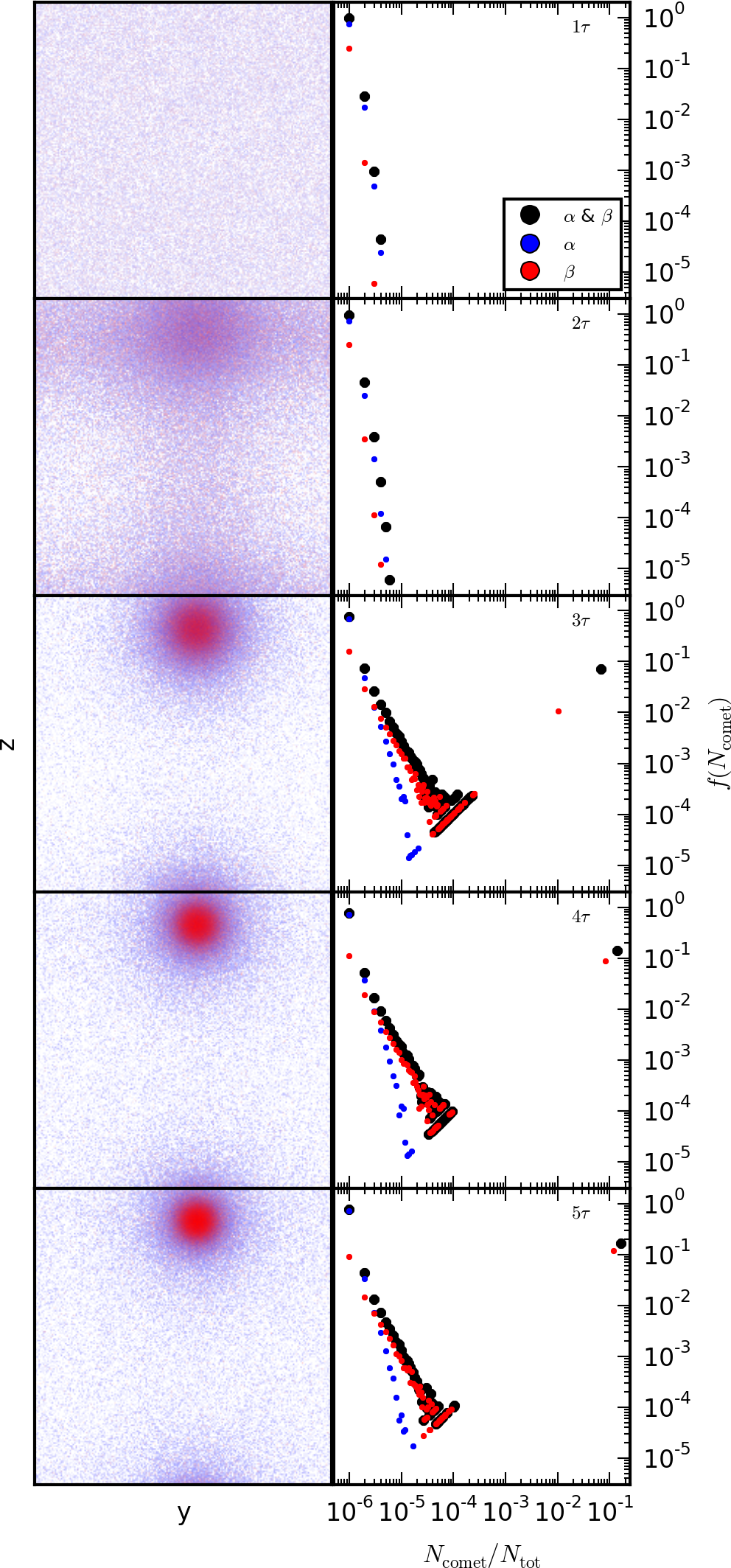}} 
        \caption{A75 with $\gamma_\name{J} = 1.5$}
        \label{fig:As}
        \end{subfigure}
        \hfill
        \begin{subfigure}[b]{0.49\textwidth}
        \centering
        {\includegraphics[width=\textwidth]{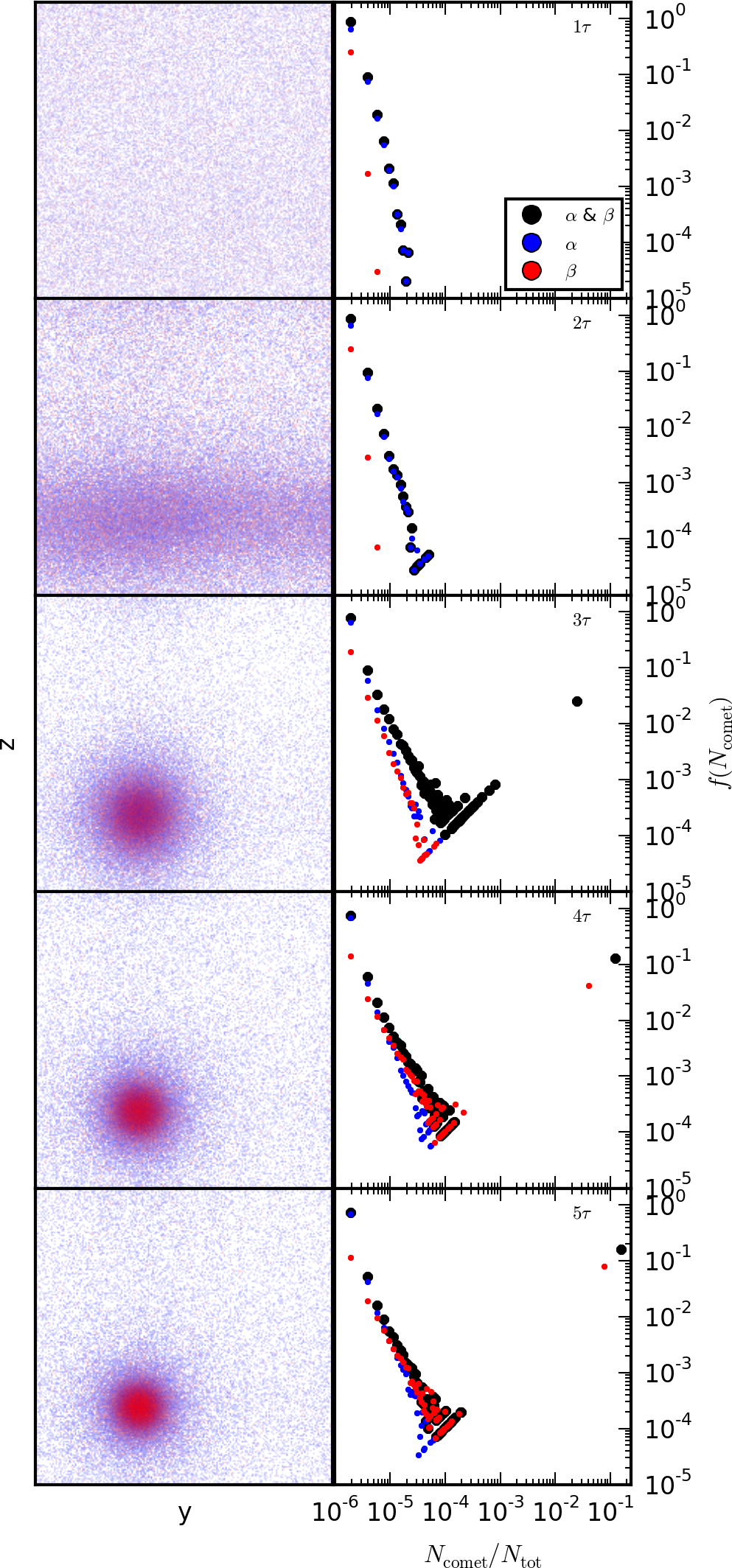}}
        \caption{B75 with $\gamma_\name{J} = 1.5$}
        \label{fig:Bs}  
   \end{subfigure}
   
        \caption{Time sequence of the simulations A75 and B75. On the left side, the slice shows in depth $20\%$ of the
            super-molecules. On the right side, $N_\name{comet}$ is the number of super-molecules in one comet and
            $f\left(N_\name{B}\right)$ is the comet size distribution function.}
\end{figure*}

\begin{figure}[t] 
        \resizebox{\hsize}{!}{\includegraphics{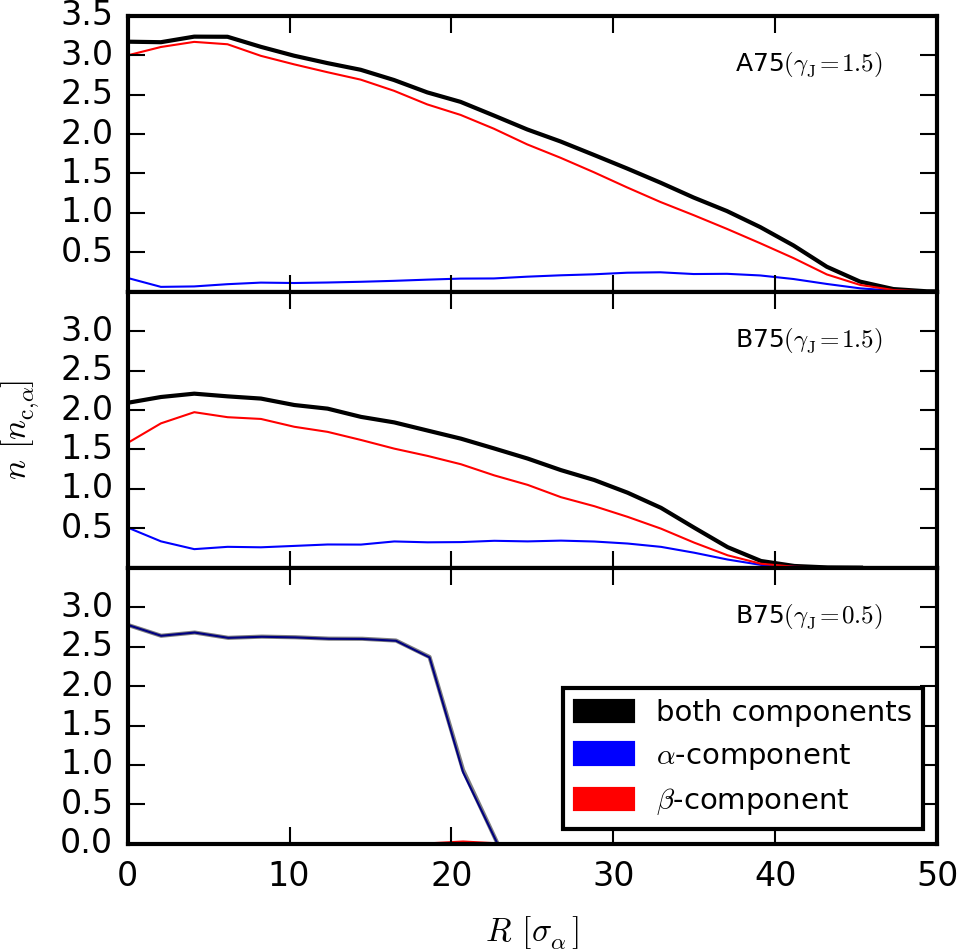}}
        \caption{Density of planetoid as a function of the radius of the simulations A75 and B75 with $\gamma_\name{J} = 1.5$
        at $t=5\tau$ and B75 with $\gamma_\name{J}=0.5$ at $t=30\tau$.}
        \label{fig:Ar}
\end{figure}

Figure \ref{fig:As} (page \pageref{fig:As}) shows a time sequence of snapshots and super-molecule, comet-size
distributions condensed as grains or comets. The parameter $N_\name{comet}$ is the number of super-molecules in one
comet and $f\left(N_\name{B}\right)$ is the comet size distribution function. At the beginning with $t<3\tau$, small
comets of either $\alpha$- or $\beta$-molecules form. At $t = 3\tau$, a planetoid with $N \approx 0.1 N_\name{tot}$
forms consisting of both components.  One can already see a dominance of $\beta$-molecules, especially in the centre.

Beginning at $t=3\tau$, and even more clearly at $t = 4\tau$, one can observe the formation of a big core consisting
only of $\beta$-molecules (isolated $\beta$-dot). In the snapshots this corresponds to the planetoid shown as a
$\beta$-core surrounded by $\alpha$-molecules. Once the planetoid has reached this form, it reaches a steady state. Its
temperature matches the gas temperature, and the temperature fluctuations level out (see FP2015 for more details on
planetoid and comet temperatures).

Figure \ref{fig:Ar} (top) shows the planetoid density of simulation A75 as a function of radius. Even though the fluid
consists of only $25\%$ $\beta$-molecules, the planetoid consists mostly of $\beta$-molecules with $f_\beta = 0.86$. 
The $\alpha$-molecules are only a small fraction and mostly present in the outer part. The gaseous nature of this body
is visible as the density regularly decreases in radii.

\subsubsection{Scaling}

\begin{figure}[t] 
        \resizebox{\hsize}{!}{\includegraphics{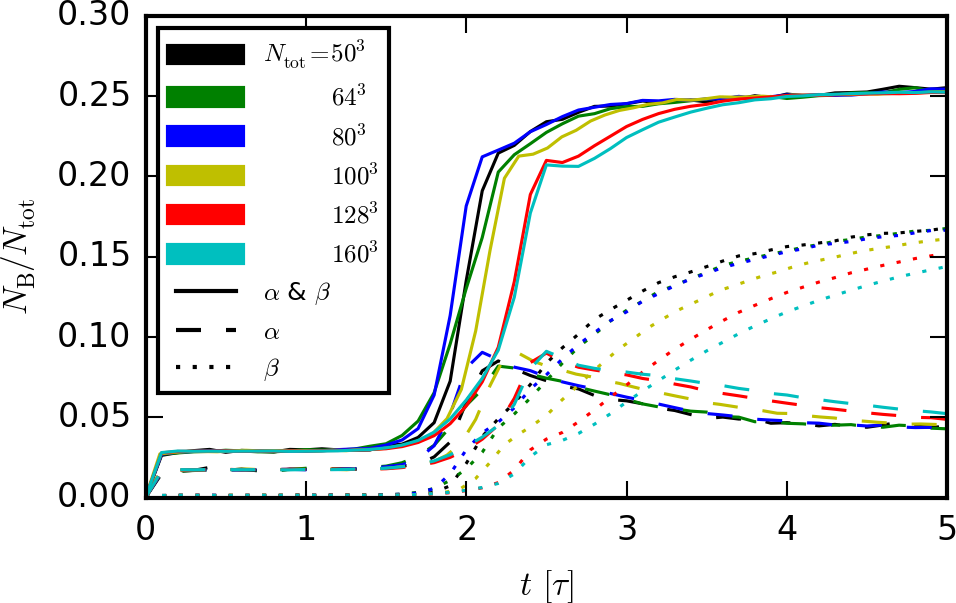}}
        \caption{Fraction of bound molecules as a function of time for the simulations A75$_\name{S}$  with  $\gamma_\name{J} =
        1.5$ and different $N_\name{tot}$.}
        \label{fig:AcScale}
\end{figure}

\begin{figure}[ht] 
        \resizebox{\hsize}{!}{\includegraphics{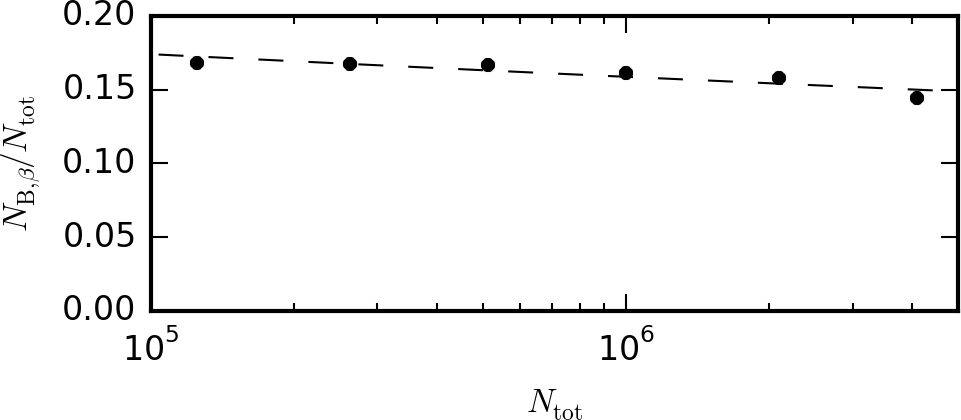}}
        \caption{Fraction of bound $\beta$-molecules as a function of $N_\name{tot}$ for the simulations A75$_\name{S}$.}
        \label{fig:AcScaleFix}
\end{figure}

The scaling of simulations using super-molecules has already been discussed in FP2015. In order to obtain the correct
behaviour,  the gravitational forces $F_\name{G}$ need to be small on intermolecular scales compared to the
Lennard-Jones forces $F_\name{LJ}$, i.e. $F_\name{G}(r_\name{c}) < F_\name{LJ}(r_\name{c}),$ where $r_\name{c}$ is the
cut-off radius (see Equ.\ \ref{equ:xglj}).  A turning point can be identified up to which
$N_\name{B}\left(N_\name{comet}\right)$ follows the power-law $N_\name{B}\left(N_\name{comet}\right) \propto
N_\name{comet}^{\xi_\name{c}}$ with negative index $\xi_\name{c}<-1$, whereas after the turning point
$N_\name{B}\left(N_\name{comet}\right)$ follows a second power law with index $\xi_\name{p}=1$ (see Fig.\ \ref{fig:As}).
The appearance time of this turning point is independent of $N_\name{tot}$, whereas the size of the comet at the turning
point scales as $N_\name{comet}/N_\name{tot} \approx 10\, N_\name{tot}^{-1}$, thus $N_\name{comet} \approx 10$. This
corresponds roughly to the smallest number of nearest neighbours in the condensed phase in 3D for which surface effects
start to be dominated by volume effects.

Figure \ref{fig:AcScale} shows the fraction of bound molecules as a function of time for all A75$_\name{S}$ simulations
with $N_\name{tot} = 50^3$ to $160^3$. As shown in FP2015, the slight time delay between the simulations can be
attributed to the random seed. In any case, the asymptotic final value is physically more important, and is the same for
all $N_\name{tot}$ when considering both components. The final value of the $\beta$-molecules, on the other hand, very
slightly declines with increasing $N_\name{tot}$ as can be seen in Fig.\ \ref{fig:AcScaleFix}. It follows the power-law
$N_\name{B}/N_\name{tot} \approx 0.213 \,N_\name{tot}^{-0.017}$ over the range $N_\name{tot} = 10^5-10^{6.5}$.

\subsubsection{Extrapolation to physical scale}

The simulations should actually represent  a H$_2$-He fluid mixture with ${\sim}10^{50}$ molecules.  As outrageous as
this extrapolation might appear, this is exactly what usually takes place in many other types of simulations
(cosmological, galactic, or stellar simulations) because as long as the physical scale invariant aspect of the physics
between the macro- and micro-scales are separated by enough orders of magnitude the exact range of scale difference does
not matter over dynamical timescales.  For longer simulation timescales one can check how the results scale with $N$ by
running simulations with different $N$, which is the reason why we always run the simulations with several $N$.
Extrapolating the previous power law to physical scales, we find that ${\sim}3\%$ of $\beta$-molecules settle inside the
planetoid, instead of ${\sim}15\%$.  Thus the simulations overestimate species segregation, which is to be expected in
view of the increased fluctuations when the number of particles decreases.  Segregation effects should be treated with
caution, as we are extrapolating values in a range that is less than two orders of magnitude or 45 orders of magnitude
away. Larger simulations should allow us to better constrain the effective species segregation in realistic conditions.

\subsection{Between critical temperatures}

In this section, we consider fluids with $T_\alpha < T_\name{c, \alpha}$ and $T_\beta > T_\name{c, \beta}$ with
different component fraction $x_\alpha$. The number density $n$ has been chosen in such a way that for the molecular
fractions $x_\alpha > 0$, the $\alpha$-component with number density $n_\alpha = x_\alpha\cdot n$ is in a phase
transition.

As the temperature of the B simulations is an order of magnitude smaller than in the A simulations, the same is the case
for the gravitational potential (see Equ.\ \ref{equ:GJ}). For that reason, it is sufficient to use $N_\name{tot} = 80^3$
for these  simulations.

\subsubsection{Above the ideal gas Jeans criterion}
\label{subsec:Bab}

\begin{figure*}[p] 
        \centering      
        \begin{subfigure}[b]{0.49\textwidth}
        \centering
        {\includegraphics[width=\textwidth]{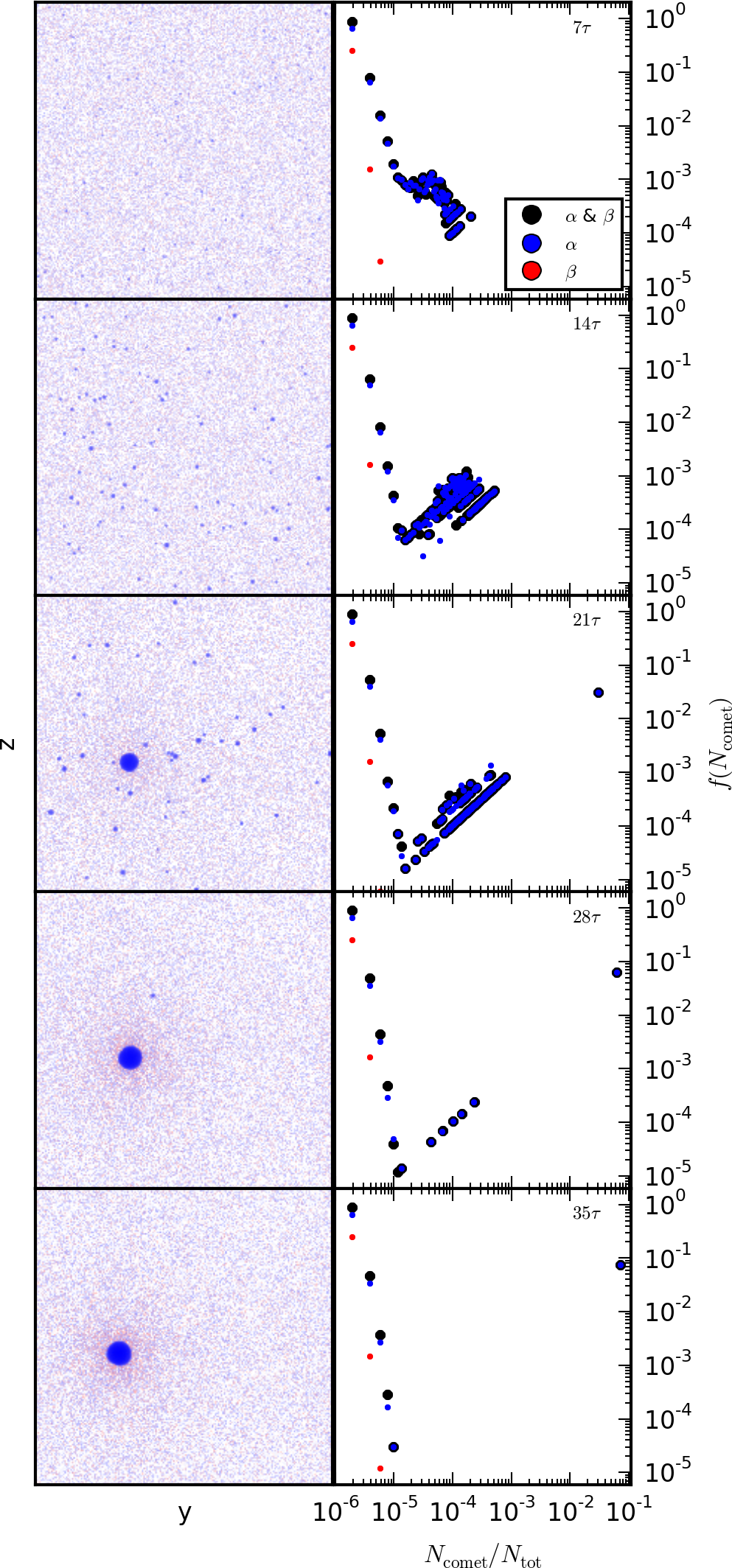}}
        \caption{B75 with $\gamma_\name{J} = 0.5$}
        \label{fig:B05s}
        \end{subfigure}
        \hfill
        \begin{subfigure}[b]{0.49\textwidth}
        \centering
        {\includegraphics[width=\textwidth]{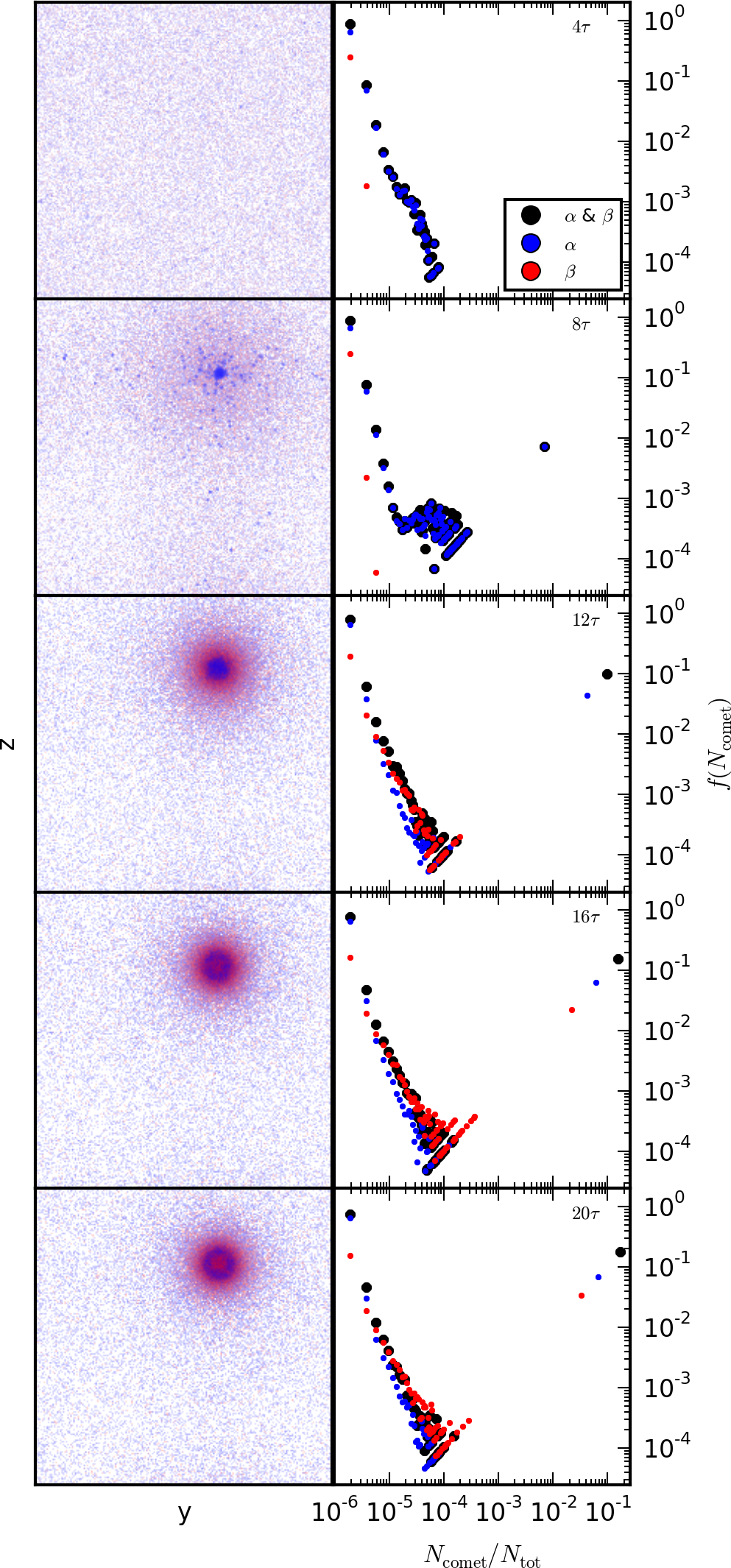}}
        \caption{B75 with $\gamma_\name{J} = 0.8$}
        \label{fig:B75G08s}
        \end{subfigure}
        
        \caption{Time sequence of the simulation B75. On the left side, the slice shows in depth $20\%$ of the
                super-molecules. On the right side, $N_\name{comet}$ is the number of super-molecules in one comet and
                $f\left(N_\name{B}\right)$ is the comet size distribution function.}

\end{figure*}

Figure \ref{fig:simNB}B on the left side shows the time evolution of the fraction of bound molecules for the B
simulations with $\gamma_\name{J} = 1.5$. One can see the similarity to Fig.\ \ref{fig:simNB}A, but the fluids with a
high $x_\alpha$ value are rising to higher values even before the perturbation is becoming dominant. This is because the
$\alpha$-portion of the fluid is in a phase transition and small ice grains are forming even without the help of
gravity. The formed planetoid is gaseous, as can be seen in  Fig.\ \ref{fig:Ar} (middle). This shows that the phase
transition does not have an important effect if $\gamma_\name{J} > 1$ and that the instability can be predicted by the
ideal gas Jeans criterion.

The density at the core of the planetoid of simulation B75 is lower than that of A75. This is explained by the fact that
by keeping $\gamma_\name{J} = 1.5$, the value for $G_\name{J}$ is lower for the B simulations than for the A simulations
as $G_\name{J} \propto T$ (see Equ.\ \ref{equ:GJ}). Having a lower gravitational potential, the density at which the
repulsive Lennard-Jones term and the attractive gravitational term are equal is lower.

The planetoid is still dominated by $\beta$-molecules, but there are more $\alpha$-molecules than in the A75 simulation
with $f_\alpha = 0.26$ and  $f_\beta = 0.74$. As in the A simulations, the core consists of gaseous He, as is clearly
visible in Fig.\ \ref{fig:Bs} (page \pageref{fig:Bs}).

\subsubsection{Below the ideal gas Jeans criterion}
\label{subsec:Bbel}

As can be seen in Fig.\ \ref{fig:cases}, the $\alpha$-component of the simulations B10, B75, B50, and B25 all lie on the
Maxwell line and are thus in a phase transition, which implies, according to Equ.\ (\ref{equ:kM}), that they are
gravitationally unstable even with $\gamma_\name{J} < 1$.

The right side of Figure \ref{fig:simNB}B shows the evolution of the fraction of bound molecules of the B simulations
with $\gamma_\name{J}=0.5$. The timescale is much larger ($30\tau$  instead of $5\tau$ in the case of $\gamma_\name{J} =
1.5$), having a smaller gravitational potential, the long-range gravitational term is lower, and therefore the creation
of any potential comet or planetoid takes more time.

The one-component fluids, consisting of either uniquely $\alpha$-molecules (B10) or $\beta$-molecules (B00) have already
been studied in detail in FP2015. The $\alpha$-fluid B10 is unstable as it is in a phase transition, whereas the
$\beta$-fluid B00 is stable as its temperature is above the critical value and no phase transition is possible.

The simulations of fluid mixtures B25, B50, B75 with $\gamma_\name{J} = 0.5$ are all unstable, even gravitationally. We
can distinguish a clear difference in the simulations with $\gamma_\name{J} = 1.5$ in that only the $\alpha$-molecules
form comets, whereas the $\beta$-molecules remain in gaseous form. Even in the simulation B25, which has only $25\%$
$\alpha$-molecules, the comets and planetoid consist almost exclusively of $\alpha$-molecules. This difference between
$\gamma_\name{J} = 1.5$ and $0.5$ can also be seen when comparing Fig.\ \ref{fig:B05s} with Fig.\ \ref{fig:Bs} (pages
\pageref{fig:B05s} and \pageref{fig:Bs}).

Figure \ref{fig:Ar} (bottom) shows the radius of the planetoid at $t=30\tau$ of B75 with $\gamma_\name{J}=0.5$.
Comparing with the planetoid of B75 with $\gamma_\name{J} = 1.5$, we see that the high-gravity planetoid consists mostly
of $\beta$-molecules in gas phase, whereas the low-gravity planetoid consists of mostly $\alpha$-molecules in solid
phase, surrounded by an atmosphere. Very few $\beta$-molecules have been trapped during the planetoid formation,
providing an interesting example of a body forming with a distinct composition from the original medium as a result of
the initial phase transition state.

\subsubsection{Different $\gamma_J$ values}

\begin{figure}[t] 
        \resizebox{\hsize}{!}{\includegraphics{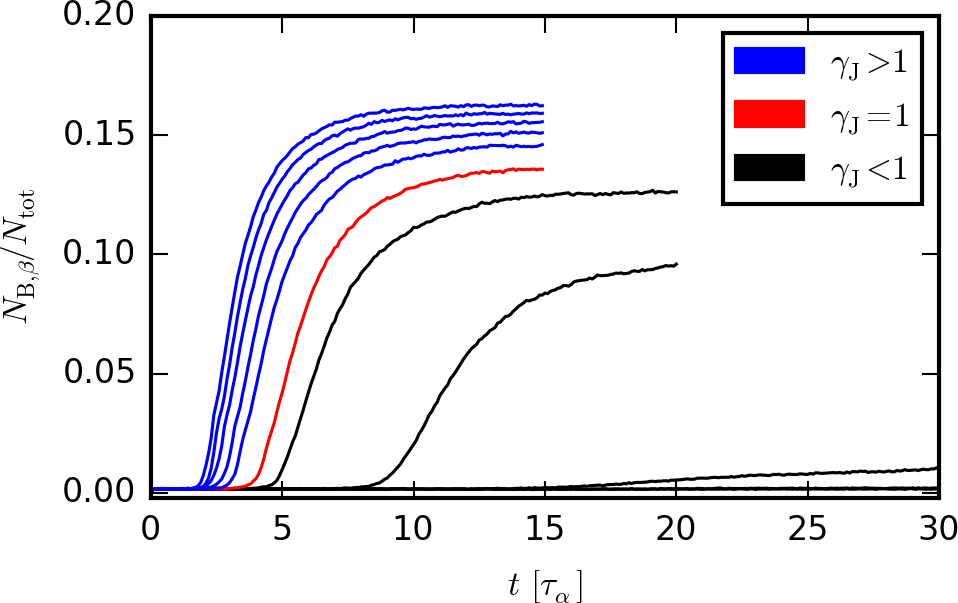}}
        \caption{Fraction of bound $\beta$-molecules as a function of time for the simulations B75$_{\gamma}$. The simulations
        are stopped once they reach asymptotic values.}
        \label{fig:BGsc}
\end{figure}

The previous sections show that a fluid in a phase transition above the ideal gas Jeans criterion, i.e. with
$\gamma_\name{J} = 1.5$, forms a gaseous planetoid consisting mostly of $\beta$-molecules due to a classical ideal gas
Jeans collapse. On the other hand, a fluid in a phase transition with $\gamma_\name{J} = 0.5$ forms small
$\alpha$-comets due to the phase transition. These comets are attracted to each other by gravity, leading to the
formation of a rocky planetoid, consisting almost exclusively of $\alpha$-molecules. In this section, we vary
$\gamma_\name{J}$ from $0.5$ to $1.5$.

Figure \ref{fig:BGsc} shows the fraction of bound $\beta$-molecules. It is rising steeply for fluids with
$\gamma_\name{J} > 1$ in accordance with the ideal gas Jeans criterion and the formed planetoid is gaseous and consists
mostly of $\beta$-molecules. The fluid with $\gamma_\name{J} = 1$ also produces a gaseous planetoid, but the percentage
of $\beta$-molecules is already dropping a little. Interestingly, in the fluids with $0.7 \leq \gamma_\name{J} < 1$, the
$\beta$ fraction is also rising. The instability criterion of Equ.\ (\ref{equ:kM}) is for all components, not only one
of them.

Figure \ref{fig:B75G08s} (page \pageref{fig:B75G08s}) shows snapshots and comet-size distributions of simulation
B75$_{\gamma}$ with $\gamma_\name{J} = 0.8$. One sees that at first ($t \leq 8\tau$) only the $\alpha$-molecules are
collapsing and forming a rocky planetoid.  Then, owing to the great attractive force of the $\alpha$-planetoid, many
$\beta$-molecules gather around it, forming an atmosphere ($t = 12\tau$). A $\beta$-atmosphere can also be observed, in
a less striking way, for the simulation with $\gamma_\name{J} = 0.5$ in Fig.\ \ref{fig:B05s}. What happens afterwards is
very interesting: at $t=16\tau$, one sees that the rocky planetoid swaps the $\alpha$- and $\beta$-molecules and the
heavier $\beta$-molecules replace the $\alpha$-molecules near the centre.

\subsection{Below critical temperatures}

In this Section, to complete the study of binary fluid mixtures, we consider fluids where both $T_\alpha$ and $T_\beta$
are below the critical temperature.  The number density $n$ has been chosen in such a way that for the molecular
fractions $x_\alpha > 0$, the $\alpha$-component with number density $n_\alpha = x_\alpha\cdot n$ is in a phase
transition.

\subsubsection{Above the ideal gas Jeans criterion}
\label{subsec:Cab}

\begin{figure*}[p] 
        {\includegraphics[width=8.5cm]{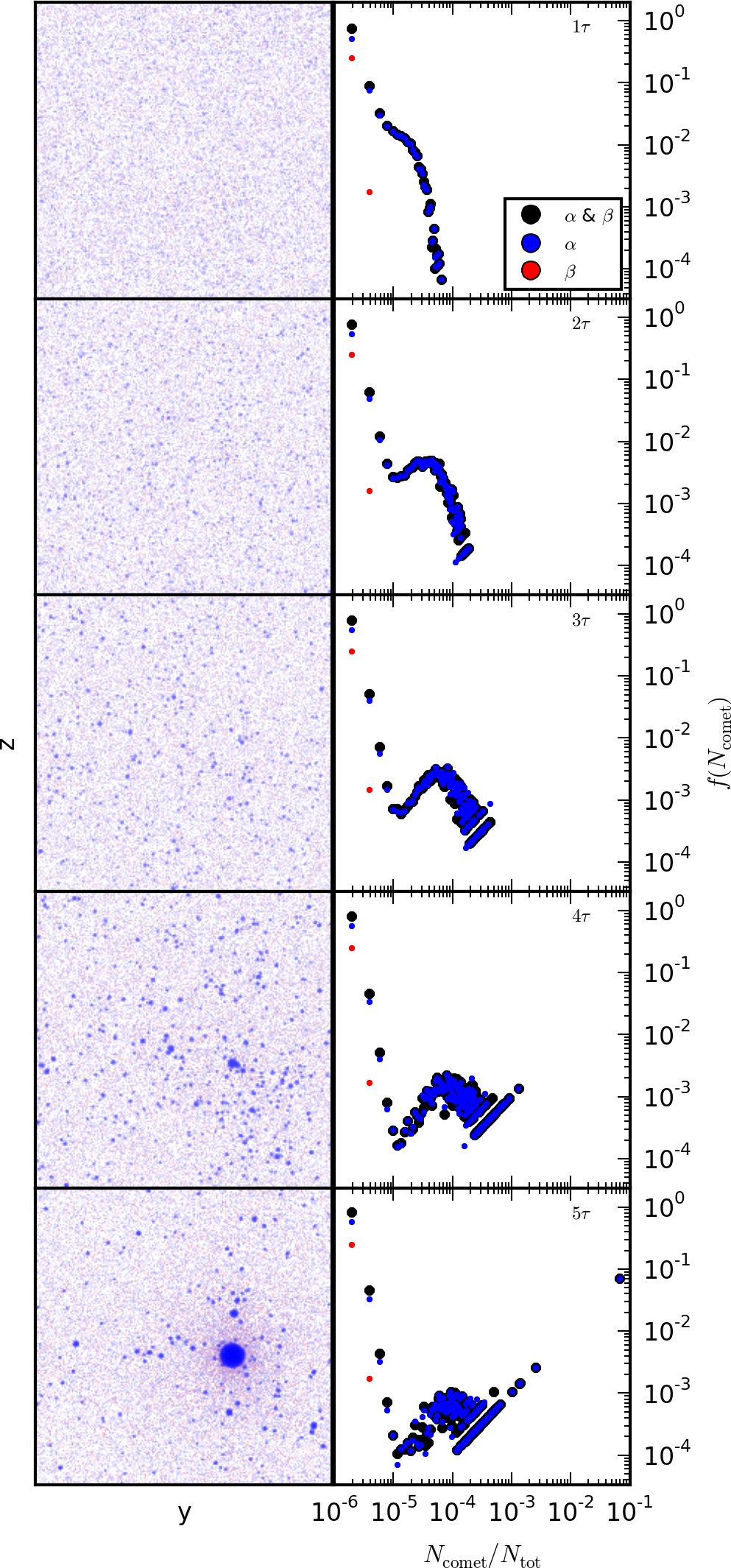}}           
        \caption{Time sequence of the simulation C75 with $\gamma_\name{G}=1.5$. On the left side, the slice shows in
            depth $20\%$ of the super-molecules. On the right side, $N_\name{comet}$ is the number of super-molecules in one
            comet and $f\left(N_\name{B}\right)$ is the comet size distribution function.}

        \label{fig:C75s}
\end{figure*}

The left side of Figure \ref{fig:simNB}C (page \pageref{fig:simNB}) shows the time evolution of bound molecules of the C
simulations with $\gamma_\name{J} = 1.5$. There is a distinct difference compared to the A and B simulations, which form
$\beta$-planetoids; only the percentage of bound $\alpha$-molecules rises and the forming planetoid only consists of
$\alpha$-molecules (see Fig.\ \ref{fig:C75s}, page \pageref{fig:C75s}). This is slightly counter-intuitive at first, as
one could expect the $\beta$-molecules to be even more eager to fall into the planetoid than in the A and B simulations,
since the temperature is lower.

Owing to the very low temperature of the C-simulation, however, the $\alpha$-molecules quickly form comets from the very
beginning. These comets are heavier than the $\beta$-molecules and decelerate faster into the planetoid.

\subsubsection{Below the ideal Gas Jeans criterion}
\label{subsec:Cbel}

The evolution of the simulations below the ideal gas Jeans criterion is analogous to the B simulations. The fraction of
bound $\alpha$-molecules in the pure $\alpha$-fluid and the mixture rise, and the fraction of bound molecules of the
pure $\beta$-fluid remains very low. This is in accordance with Fig.\ \ref{fig:cases} where the $\alpha$-molecules are
unstable but the $\beta$-molecules are stable. The simulations C10, C75, C50, and C25 form a rocky $\alpha$-planetoid,
as already seen in the B simulations (see Fig.\ \ref{fig:B05s}).

\subsection{Virial theorem} 

\begin{figure}[t] 
        \resizebox{\hsize}{!}{\includegraphics{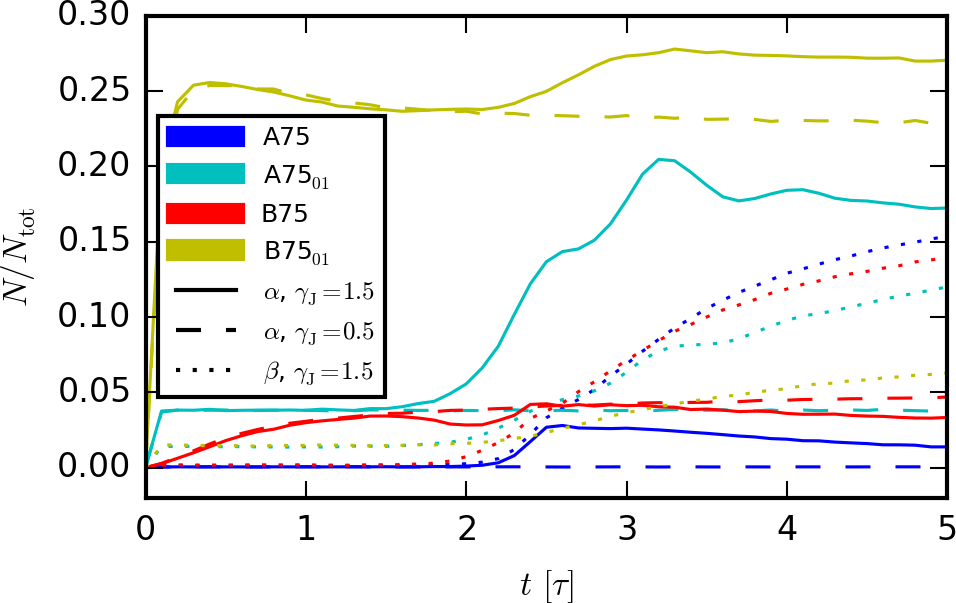}}
        \caption{Fraction of bound molecules with cluster mass $m_\name{cl} > m_\name{He}$ as a function of time of simulations
          in and out of the unvirializable density domain.}
        \label{fig:virc}
\end{figure}

\begin{figure}[t] 
        \resizebox{\hsize}{!}{\includegraphics{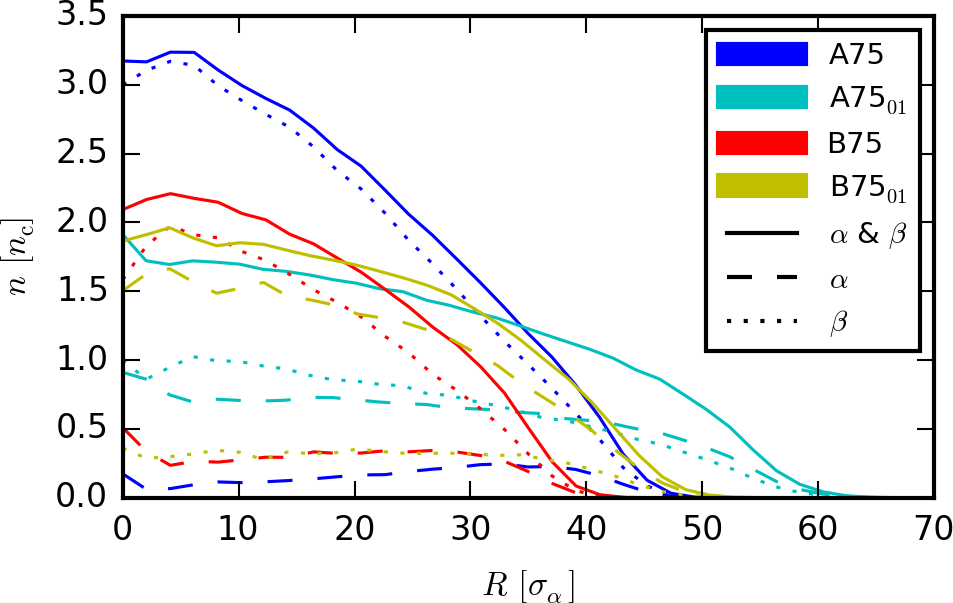}} 
        \caption{Density of planetoid  at $t=5\tau$ as a function of the radius of the simulations
        in and out of the unvirializable density domain.}
        \label{fig:virr}
\end{figure}

When comparing the simulations above the ideal gas Jeans instability, there is a clear difference between the A and B
simulations on one side, and the C simulations on the other. A gaseous $\beta$-planetoid forms in the first two, whereas
a rocky $\alpha$-planetoid forms in the latter. Looking at the virial terms of the fluids (see Sec.
\ref{subsubsect:vir}), Equ.\ (\ref{equ:AR}) is fulfilled in the A and B simulations, whereas for the C simulation, the
density is in the unvirializable domain $\mathcal{D}$. In this Section, we vary the densities of the A and B simulations
in order to be in and out of the unvirializable domain.

Figure \ref{fig:virc} shows the time evolution of clusters that have a higher mass than one $\beta$-molecule
($N_\name{cl,\alpha}\!>\!2$) for the simulations in the unvirializable domain (A75$_{01}$ and B75$_{01}$) and below (A75
and B75).  A very quick rise of H$_2$ comets for the unvirializable fluid happens, both with and without gravity, which
is in accordance with Equ.\ (\ref{equ:AR}), as neither the repulsive Lennard-Jones term nor the kinetic energy can
withhold the attractive Lennard-Jones term thus leading to the formation of comets. Even in simulation A75$_{01}$, with
a temperature above the critical temperature, this comet formation is taking place, even though a phase transition is
officially not possible. A slow comet formation only takes place for the virializable fluids.

Once the exponential growth of the perturbation becomes important ($t \geq 2\tau$), the unvirializable fluids have
created an important number of comets heavier than the $\beta$-molecules, which fall faster in the forming planetoid as
a result of dynamical friction. This can be seen in Fig.\ \ref{fig:virr} where the planetoids of the simulations A75 and
B75 consist mostly of $\beta$-molecules, whereas the planetoid of B75$_{01}$ consists mostly of $\alpha$-molecules. A
somewhat special case is A75$_{01}$, where the planetoids composition is almost perfectly fifty-fifty. This can be
explained by the fact that because is is above the critical temperature, the comets are not really solid, but consist of
a dense gas that is able to mix easily with $\beta$-molecules. Thus, once a $\alpha$-planetoid has formed using all the
heavy $\alpha$-comets, the $\beta$-comets fall into the planetoid and mix with it.

\subsection{Influence of $\beta$-molecules on $\alpha$-molecules below the ideal gas Jeans criterion}

\begin{figure}[t] 
        \resizebox{\hsize}{!}{\includegraphics{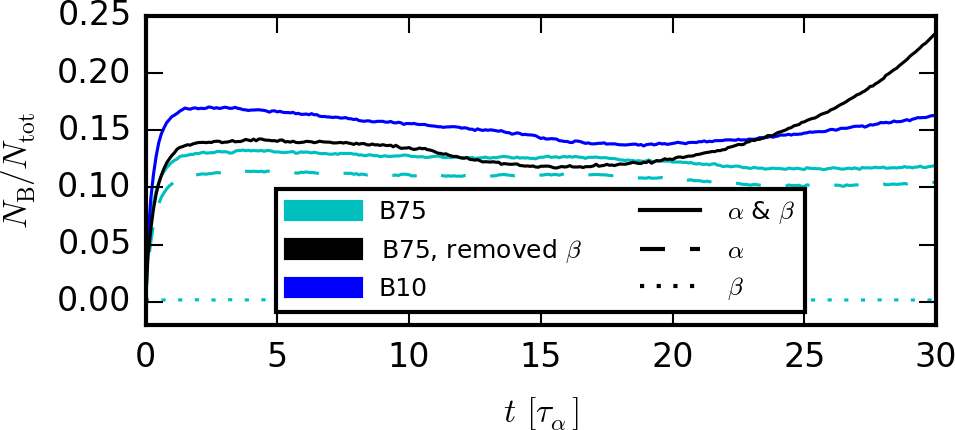}}
        \caption{Fraction of bound molecules as a function of time of the simulations B75, B75 with removed $\beta$-molecules 
                and B10. $\gamma_\name{J} = 0.5$.}
    \label{fig:B75c}
\end{figure}

As can be seen in Fig.\ \ref{fig:simNB}C, almost no $\beta$-molecules form comets if $\gamma_\name{J} = 0.5$ and the
percentage of $\beta$-molecules in the planetoid is negligible. Granted, the concentration of He around the planetoid
rises slightly as can be barely seen in Fig.\ \ref{fig:B05s}.  Thus the question can be raised whether a small fraction
of a secondary molecule (such as He in the case of molecular clouds) needs to be included in low-gravity simulations. To
answer that question, simulation B75 with $\gamma_\name{J} = 0.5$ was run again, but all $\beta$-molecules were removed
and their mass was equally distributed to the $\alpha$-molecules to maintain the same gravitational potential.

Figures \ref{fig:B75c} shows the time evolution of the fraction of bound molecules of the simulations B75, B75 without
$\beta$-molecules and B10 for comparison. Even though the two B75 simulations are similar, there are differences to be
observed. The fraction of bound $\alpha$-molecules of the simulation B75 should correspond to the total fraction of
bound molecules of the simulation without $\beta$-molecules, but the latter is higher; the $\beta$-molecules in B75 have
a damping effect on the comet formation. In addition, the simulation without $\beta$-molecules is rising to a higher
value at the end of the simulation.

The inclusion of a small fraction of a secondary molecule does change the look of the simulation by damping the comet
formation of $\alpha$-molecules. For that reason, the inclusion of secondary molecules in more realistic simulations is
useful.

\subsection{Physical systems}

\begin{figure}[t] 
        \resizebox{\hsize}{!}{\includegraphics{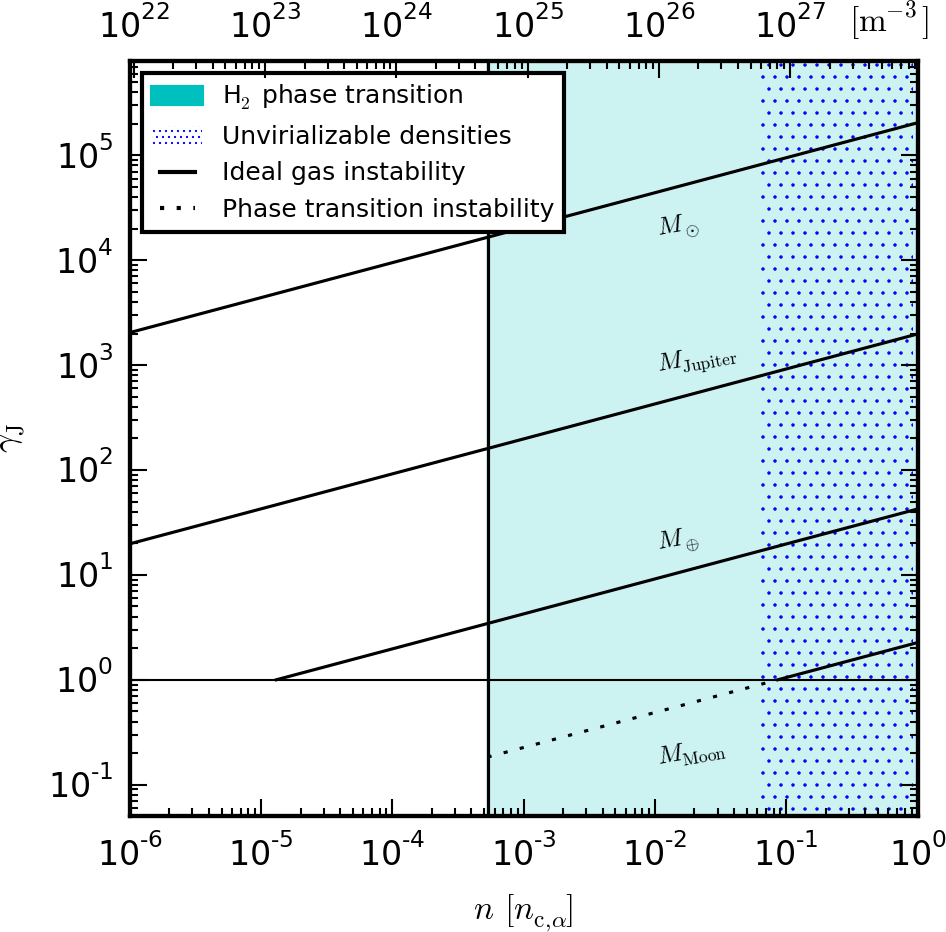}}
        \caption{$\gamma_\name{J}$ of different total fluid masses at $T=10\,\unit{K}$, as a function of number
          density, indicating either ideal gas Jeans instability, or instability owing to phase transition.}
        \label{fig:SS}
\end{figure}

\begin{figure}[t] 
        \resizebox{\hsize}{!}{\includegraphics{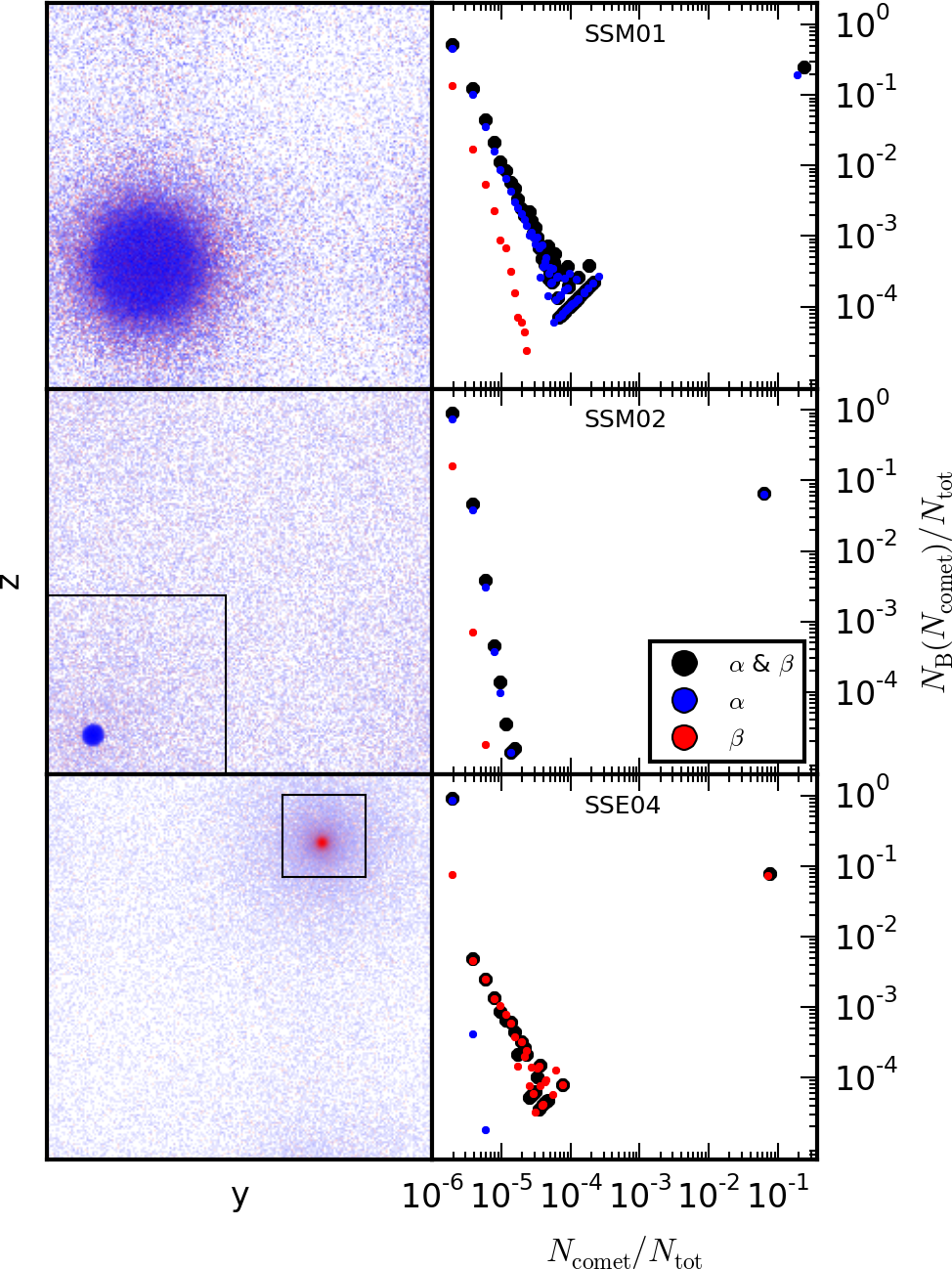}}
        \caption{Snapshots and comet-size distributions of the simulations SSM01, SSM02, and SSE04. The slice selects
          in depth $20\%$ of the super-molecules. The squares in the two lower left frames are the same size as the
          next upper frame.}
        \label{fig:SSs}
\end{figure}

Up to now, we have looked at theoretical models, varying $x_\alpha$ from $0$ to $1$, and setting the temperature and
density as a fraction of the respective $\alpha$ critical values. The critical values for H$_2$ are $T_\name{c} =
32.97\,\unit{K}$ and $n_\name{c} = 9.34\cdot 10^{27}\,\unit{m}^{-3}$. In astrophysical conditions, the He mass fraction
is between $w_\name{He, SS} = 0.2741$ for the solar system \citep{lodders_solar_2003} and $w_\name{He, MW} = 0.2486$ for
the initial Big Bang mixture \citep{cyburt_update_2008}, which translates to number fractions $x_\name{He, SS} = 0.1598$
and $x_\name{He, MW} = 0.1428$.

Figure \ref{fig:SS} shows $\gamma_\name{J}$ as a function of the number density for solar system abundances ($x = 0.16$)
and $T=10\,\unit{K}$ with total masses equal to the Moon, Earth, Jupiter, and Sun. H$_2$ is then in a phase transition
for $n > 4\cdot 10^{24}\,\unit{m^{-3}}$; only a Moon mass or below can be in a phase transition and below the Jeans
criterion. The fluid is unvirializable for $n > 6\cdot 10^{26}\,\unit{m^{-3}}$.

If we go to a lower temperature, say the CMB $2.7\,\unit{K}$, a H$_2$ phase transition takes place for $n >
10^{12}\,\unit{m}^{-3}$. In that case, fluids with Earth mass would be chemically unstable below the ideal gas Jeans
criterion for $n\leq 3\cdot 10^{21}\,\unit{m}^{-3}$ and with Jupiter mass for $n \leq 3\cdot 10^{16}\,\unit{m}^{-3}$.
Fluids with Sun mass, on the other hand, cross $\gamma_\name{J} = 1$ only in the gaseous phase of H$_2$. The lowest
unvirializable density $n_- = 6\cdot 10^{26}\,\unit{m}^{-3}$ does not change a lot with temperature.

The number of FFT mesh cells $N_\name{FFT} \propto L^3$ and the simulation timescale $\tau \propto L$ both directly
depend on $L \propto n^{-1/3}$, and the total calculation duration scales as $t_\name{sim} \propto \tau\cdot
N_\name{FFT} \propto n^{-4/3}$. For that reason, simulating a fluid at CMB temperature with densities below
$10^{20}\,\unit{m}^{-3}$ would translate to extremely long simulation run times with today's computers. In addition, the
upper limit for the mass of super-molecules is $m_\name{SM, max} \approx 5\cdot 10^{-6} M_\oplus$ (see Equ.\
\ref{equ:xglj}). Thus, the minimum number of super-molecules $N_\name{tot,\, min} = M/m_\name{SM, max}$ is ${\sim}2\cdot
10^5$, $6.5\cdot 10^7$, $6.5\cdot 10^{10}$ for simulating an Earth,  Jupiter, and Sun mass, respectively. For that
reason, for the time being we content ourselves to studying systems up to total mass comparable to the Earth mass.

\subsubsection{Planetoid formation}

Three simulations were run at a temperature of $T=10\,\unit{K}$, which is above the critical temperature of He and below
that of H$_2$, and thus in a similar regime as the B simulations. Two simulations have a total mass equal to the Moon,
with $n \approx 10^{27}\,\unit{m}^{-3}$ which is above the ideal gas Jeans criterion and in the unvirializable domain,
and $n \approx 10^{26}\,\unit{m}^{-3}$, which is below the Jeans criterion, and one has a total mass equal to the Earth
and with $n \approx 10^{24}\,\unit{m}^{-3}$, which is above the criterion. The simulation parameters are given in Table
\ref{tab:aT}.

Figure \ref{fig:SSs} shows the snapshot and comet-size distribution of the three simulations after the formation of a
planetoid. The fluid of SSE04 is above the ideal gas Jeans criterion and we observe the formation a He-planetoid,
surrounded by H$_2$, similar to Sect.\ \ref{subsec:Bab}. The evolution of simulation SSM02, which is below the ideal gas
Jeans criterion, leads to the formation of a rocky H$_2$ planetoid, similar to Sect.\ \ref{subsec:Bbel}.

In the case of SSM01, the density lies in the unvirializable domain, resulting in a formation of many H$_2$-grains that
are heavier than the He-atoms from the very beginning. This leads to the formation of a H$_2$-planetoid similar to
Sect.\ \ref{subsec:Cab}.

\section{Conclusions} 

In our first article, FP2015, we studied the gravitational instability of a fluid in a phase transition. We extrapolated
the results to the ubiquitous H$_2$ and showed that the formation of cold, mostly undetectable comet- and even
planet-sized rocky H$_2$ clumps is very possible. The use of only one component gives a good first impression, but in
cosmic gases, there is a mass fraction of $w \approx 25 \pm 2\,\%$ He atoms.

In the present work, we studied binary fluid mixtures analytically and via numerical simulations. The results show that,
depending on the circumstances, either  He or H$_2$ planetoids can form.

\subsection{Analytic results}

The stability of a multicomponent fluid mixture has already been studied in the literature, mostly to study fluid
binaries consisting of baryonic and dark matter. The wave number below which a fluid mixture is unstable is the sum of
the Jeans wave-numbers of each component. Since the Jeans wave number is inversely proportional to $(\partial P /
\partial \rho)^{-1}_s$, which is equal to zero in the case of a phase transition, a fluid mixture is unstable as soon as
one of its components is in a phase transition.  Physically what happens is that when one species is in a phase
transition, an overdensity only increases its condensed phase fraction at constant pressure, instead of increasing
pressure and producing no global force to counter gravity.  The transformation from the gas to the condensed phase
continues until the species is fully condensed.

We studied the evolution of unstable fluid mixtures  with the widely used Lennard-Jones intermolecular potential, which
reproduces the H$_2$ phase transition very well (but it reproduces  the He transition, which is not essential in this
work, less well). We showed, using the virial analysis of Lennard-Jones fluid mixtures, that there is a unvirializable
density-domain $\mathcal{D}$ within which the attractive forces dominate the repulsive forces for any total mass $M$ and
no virial equilibrium is possible. These states can be reached in strongly dynamical situations (e.g. during collapses)
and are able to produce condensed comets particularly quickly.  Dynamical friction is important to separate species and
condensed comets.  For instance, if H$_2$ is in a phase transition, the formed H$_2$ comets are heavier than the
He-molecules, and precipitate in a gravitational field, producing almost pure H$_2$ bodies.

There are three reasons to concentrate on plane-parallel initial collapses, as described in more detail in App. B:

\begin{enumerate}
\item In typical cosmic conditions, the fastest collapsing geometry is sheet-like, not filament- or point-like.
\item The adiabatic matter compression during collapse leads to the least heating in sheet-like geometry: in a sheet-like
  adiabatic collapse the gravitational energy released to the fluid is \textit{finite} and amounts to a maximum increase
  of temperature by only a factor of about two, while in filament-like collapses the temperature diverges
  logarithmically as a function of filament radius, and in point-like collapses the temperature diverges as the inverse
  sphere radius.
\item Radiative cooling is the easiest in sheet-like collapse. Indeed the absorption probability in sheet-like geometry
  remains almost unchanged for any compression, and an initially transparent medium remains transparent, whereas the
  probability converges to one in filament-like and point-like geometries.  Therefore, radiative cooling is barely
  slowed down in sheet-like collapses and, unlike in spherical or filament collapses, opacity is unable to prevent
  density from reaching high values.  This is a crucial point for this study, as the ISM conditions are commonly thought
  to be far away from the H$_2$ phase transition conditions.
\end{enumerate}

\subsection{Simulations}

As in FP2015, we used super-molecules to combine the Lennard-Jones intermolecular potential together with the
gravitational potential in numerical simulations. Several binary fluid mixtures were studied using two components:
$\alpha$ and $\beta$. Their respective properties (the most important being $m_\alpha/m_\beta < 1$ and
$T_\name{c,\alpha}/T_\name{c,\beta} > 1$) were chosen to mimic a H$_2$-He fluid, but the general properties of the
fluids were made molecule independent.

Three temperature domains can be defined: (A) above both critical temperatures, (B) between the critical temperatures,
and (C) below both critical temperatures. In all three cases, the molecular fraction was varied and the fluids were
simulated above and below the Jeans criterion. We used different numbers of molecules to test the scaling of the
simulations.   The precise number of super-molecules is not important for dynamical processes, but we found a weak
dependence for segregation effects in the sense that coarser simulations exaggerate these effects.

In case (A), both components are gaseous and an introduced perturbation does not grow when the gravitational potential
is below the Jeans criterion. When above the Jeans criterion, the fluid collapses and forms a gaseous planetoid.  The
$\beta$-molecules are twice as massive as the $\alpha$-molecules, and fall faster into the planetoid. For that reason,
the planetoid consists mostly of $\beta$-molecules, surrounded by an $\alpha$-atmosphere. This is independent of the
molecular fraction $x_\alpha$, even at very high $x_\alpha$-values, the planetoids consists mostly of $\beta$-molecules.

In case (B) and (C), the number density of the fluids was chosen so that the $\alpha$-component is in a phase transition
for all $x_\alpha > 0$. In fact, both cases are very similar since in both cases the $\beta$-component is not in a phase
transition. When the fluids are below the Jeans criterion, an instability happens because of the phase transition of the
$\alpha$-component, which leads to the formation of H$_2$ comets and ultimately a rocky $\alpha$-planetoid. This
planetoid is surrounded by a $\beta$-atmosphere, which is getting more important with increasing gravitational
potential. As in case (A), the molecular fraction $x_\alpha$ does not matter, even at very low $x_\alpha$-values, the
planetoid still consists almost exclusively of $\alpha$-molecules.

A suprising observation occurs for cases (B) or (C) above the Jeans criterion. In that case, there is a race between the
formation of small $\alpha$-grains owing to the phase transition and the exponential growth of the perturbation.  The
heaviest bodies are decelerated faster and fall  into the forming planetoid first. When the $\alpha$-component is either
gaseous or only forming very few and small comets, a $\beta$-planetoid forms. On the other hand, if the
$\alpha$-component forms many grains that are heavier than the $\beta$-molecules, an $\alpha$-planetoid forms. We showed
in the simulations that this race between $\alpha$ and $\beta$ is linked with the unvirializable density domain
$\mathcal{D}$. If a fluid reaches this domain, the $\alpha$-component wins, otherwise the $\beta$-component wins.

\subsubsection{Solar system abundances}

In addition to the above-mentioned simulations, fluids with solar system abundances and Moon or Earth mass were
simulated. As shown in Fig.\ \ref{fig:SS}, a fluid with Earth mass cannot be below the Jeans criterion and still in a
phase transition, but with Moon mass, this is possible. In that case, a rocky H$_2$ planetoid results.  With a mass as
low as the Moon, the fluid needs to be very dense to be above the Jeans criterion. In fact, the fluid would lie in the
unvirializable density-domain $\mathcal{D}$ and, thereby, a H$_2$-planetoid forms. For a fluid with Earth mass, on the
other hand, even a relatively low-density fluid is still above the Jeans criterion. The result is a gaseous He planetoid
with a H$_2$ atmosphere.

\subsection{Instability in H$_2$-He fluid}

\begin{figure}[t] 
        \resizebox{\hsize}{!}{\includegraphics{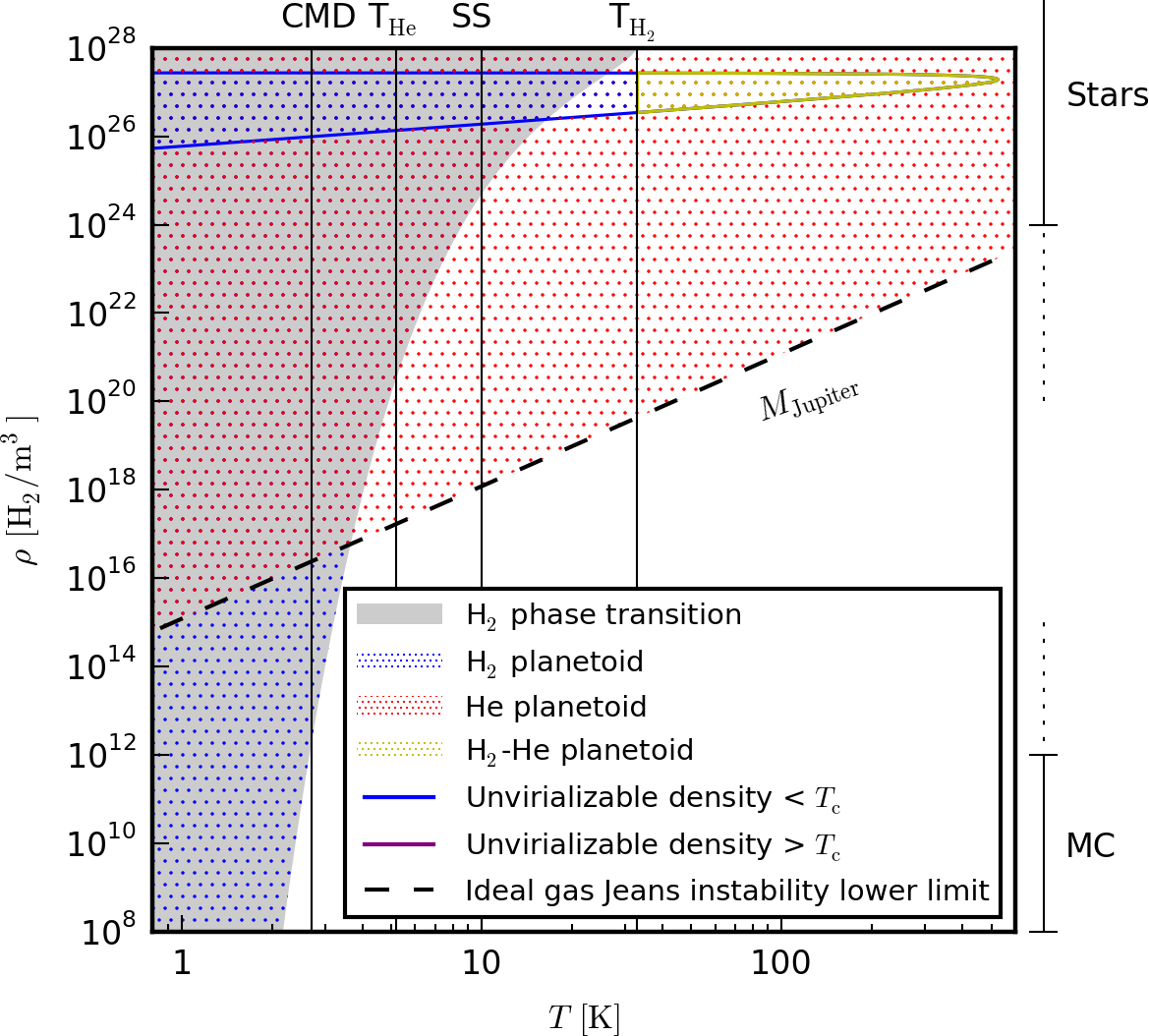}} 
        \caption[]{Gravitational instabilities at different temperatures and densities for a fluid with Jupiter mass.}
        \label{fig:domains}
\end{figure}

Figure \ref{fig:domains} shows different possible planetoid and comet formations due to gravitational instability for a
fluid with Jupiter mass. A fluid is gaseous if it is below the phase transition domain and a fluid is solid or liquid if
above. When the density is in the phase transition, it can rise without an increase of pressure.

There can be no formation below the Jeans criterion if the fluid is not in a phase transition. Most of the planetoids
due to an ideal gas Jeans collapse consist of gaseous He, but if the fluid is in the unvirializable domain
$\mathcal{D}$, then a H$_2$ planetoid forms. This H$_2$ planetoid can be solid/liquid or gaseous depending on its
temperature. If gaseous, He is able to percolate down, slowly transforming it into a He planetoid.

If the fluid is in a phase transition, we have to distinguish between a collapse above the ideal Jeans criterion, which
leads to a gaseous He planetoid except in the unvirializable domain, where it becomes a rocky H$_2$ planetoid, and in a
collapse below the ideal Jeans criterion, which also leads to a rocky H$_2$ planetoid.

The usual average density domain of molecular clouds lies between $10^8$ and $10^{12}\,\unit{H_2/m^3}$ and, with such a
density, a H$_2$ phase transition is only possible at temperatures below ${\sim}5\,$K.  However, molecular clouds are
observed to follow a fractal mass distribution over a minimum of 4--6 orders of magnitude in column densities, so the
average density is not a quantity to characterize molecular clouds properly.  Since we know that stars form with
densities ${\sim}10^{29}\,\unit{H_2/m^3}$, by continuing this argument, intermediate states covering all this density
interval have to exist.

Fluids with a high total mass, especially with stellar mass or above, reach the ideal gas Jeans criterion very quickly
leading to gaseous He-planetoids. Fluids with lower total mass, however, as for example the cold globules observed in
the Helix nebula, especially with Earth mass and below, have the ideal gas Jeans criterion at much higher densities and
are in the phase transition domain before being above the ideal gas Jeans criterion.

\subsection{Perspectives}

This and the previous FP2015 study show that the cold ISM physics is much richer than previously imagined.  The
formation of substellar gaseous or rocky condensed bodies by the H$_2$ phase transition combined to gravity, appears
natural once we recognize that collapses proceed most of the time along the sequence pancake, filament, and point, and
in the first sheet-like phase high densities allowing a H$_2$ phase transition can be reached if the initial medium
temperature is below ${\sim}15\,$K. This temperature limit would be even higher if radiative cooling had been
considered.  In the isothermal case this limit is ${\sim}33\,$K.

Most of the ISM cold gas must therefore pass over molecular cloud lifetimes (${\sim}10^6-10^8\,$yr) through such brief
(${\sim}10^2-10^4\,$yr) singular sheet-like collapses where density diverges but not temperature.  Observationally, such
events are difficult to detect because of the limited increase of temperature, opacity, and column density all along the
collapse, while reaching high volume densities.  When seen edge-on such sheet-like collapses would look like filaments.

The simulations we were able to perform are still very limited in total mass.  Including He is necessary but this
provides a number of complications with respect to the pure H$_2$ case, and widens the general picture found in FP2015.
Combining the accumulated experience of large-scale gas phase simulations by other authors
\citep[e.g.][]{renaud_sub-parsec_2013,butler_kiloparsec-scale_2015}, we can easily extrapolate what larger simulations
should produce with micro-AU resolution. Instead of one planetoid per simulation box, pc-sized sheet-like collapses
should show filaments with longer lifetimes, which would funnel H$_2$ condensed bodies and produce a spectrum of
planetoids,  comets, and occasionally stars.  The leftover condensed cold substellar bodies should then start to
evaporate according to the ambient radiation flux and depth of their gravitational potential.  The lifetime of such
bodies should be short near the centre of galaxies, but much longer at the periphery of galaxies, or even in
intergalactic space, especially in cosmic filaments.  One can postulate that, especially at the periphery of disk
galaxies where the radiation heating is low, some fraction of the dark baryons can be trapped in the form of such
condensed bodies. We plan to pursue further simulation work to deepen our understanding of the processes associating
phase transition with gravitational dynamics.

\begin{acknowledgements}
  This work is supported by the STARFORM Sinergia Project funded by the Swiss National Science Foundation.  We thank the
  LAMMPS team for providing a powerful open source tool to the scientific community.
\end{acknowledgements}

\bibliographystyle{aa} 
\bibliography{FP2016} 
\addcontentsline{toc}{section}{\refname}

\newpage
\begin{appendix}
        
\section{Jeans instability}

We first recall the classical Jeans criterion for a one-component fluid, and then we show how the same approach can be
used to find the solution of a two-component fluid. See \citet{grishchuk_gravitational_1981} for the solution of an
$n$-component fluid.

\subsection{One component}
\label{app:oc}

The equations for conservation of mass and momentum and for the gravitational potential of a fluid are written as
\begin{eqnarray}
{\partial \rho \over \partial t} + \nabla\cdot(\rho \vec{v}) &=& 0 \ ,  \\
{\partial  \rho\vec{v} \over \partial t} + \nabla\cdot\left(\rho \vec{v}\vec{v}\right) 
+ \nabla P  +  \rho\nabla\Phi &=& 0 \ , \\
\nabla \cdot \nabla\Phi - 4\pi G\rho&=& 0 \ . 
\end{eqnarray}
Following \citet{jeans_stability_1902}, we supersede these equations with perturbation terms in the $x$ direction $\rho
= \rho_0 + \delta \rho$, $P = P_0 + \delta P$, $\Phi = \Phi_0 + \delta \Phi$ and $\vec{v} = \vec{v}_0 + \delta \vec{v}$
with $\delta\vec{v} = (\delta v, 0, 0)^T$, linearizing the equations and setting $\delta P = \left({\partial P\over
    \partial \rho}\right)_s\delta \rho$, i.e.
\begin{eqnarray}
{\partial\,\delta\rho \over \partial t} + \rho_0\nabla\cdot\delta v &=& 0 \ , \label{equ:cMass}\\
\rho_0{\partial\,\delta v \over \partial t} + \left({\partial P\over \partial \rho}\right)_s \nabla \delta \rho  + \rho_0\nabla\delta \Phi&=& 0 \ ,  \label{equ:cMom} \\
\nabla \cdot \nabla\delta \Phi - 4\pi G \delta \rho &=&  0\ . \label{equ:lin}
\end{eqnarray}
This system of partial differential equations is transformed to an algebraic system of linear equations in the Fourier
space: $\delta A = \int \!\dd k \, \hat{A}(k) \exp[i(k x - \omega t)]$, where $A$ represents $\rho$, $v$, and $\Phi$.
The passage to Fourier space transforms the differential operators $\partial / \partial t$ and $\partial / \partial x$
to multiplications by  $ -i\omega$ and $ik$, respectively,
\begin{eqnarray}
-i\omega\,\hat{\rho} + i k\,\rho_0 \hat{v} &=& 0 \ , \\
-i\omega\, \rho_0 \hat{v} +  i k\,\left({\partial P\over \partial \rho}\right)_s \hat{\rho} 
+ i k \,\rho_0 \hat{\Phi} &=& 0 \ ,\\
-k^2\hat{\Phi} - 4\pi G \hat{\rho}&=& 0 \ ,
\end{eqnarray}
which can be written in matrix form $\matr{A}\cdot\vec{x} = \vec{0}$,
\begin{equation}
\begin{bmatrix}
\omega & k\rho_0 & 0\\
k\,({\partial P / \partial \rho})_s & \omega\, \rho_0 & k \,\rho_0 \\
- 4\pi G& 0& -k^2
\end{bmatrix} \cdot 
\begin{bmatrix}
\hat{\rho} \\ \hat{v} \\ \hat{\Phi}
\end{bmatrix}
= \begin{bmatrix}
0 \\ 0 \\ 0
\end{bmatrix} \ .
\end{equation}
Non-trivial solutions for $\vec{x}$ require that the determinant of $\matr{A}$ vanishes,
\begin{equation}
\det(\matr{A}) = k^2\rho_0 \left[\omega^2 + 4\pi G\rho_0 - k^2 \left({\partial P\over \partial \rho}\right)_s \right] \ ,
\end{equation}
which is the case for either $k=0$, or
\begin{equation}
\omega^2 = k^2 \left({\partial P\over \partial \rho}\right)_s - 4\pi G\rho_0 \ .
\end{equation}
A fluid is unstable if $\omega^2 < 0$, which is the case if
\begin{equation}
k^2 < k^2_\name{J} \equiv {4\pi G\rho_0 \over (\partial P/ \partial \rho)_s} \ .
\end{equation}

\subsection{Two components}	
\label{app:tc}

Having two components $\alpha$ and $\beta$, the mass and momentum conservation have to be fulfilled for each component
as follows:
\begin{eqnarray}
{\partial \rho_{\alpha} \over \partial t} + \nabla\cdot(\rho_{\alpha} \vec{v}_{\alpha}) &=& 0 \ ,  \\
{\partial \rho_{\beta} \over \partial t} + \nabla\cdot(\rho_{\beta} \vec{v}_{\beta}) &=& 0 \ ,  \\
{\partial \rho_{\alpha}\vec{v}_{\alpha} \over \partial t} + \nabla\cdot\left(\rho_{\alpha} \vec{v}_{\alpha}\vec{v}_{\alpha}\right) 
+ \nabla P_{\alpha}  +  \rho\nabla\Phi &=& 0 \ , \\
{\partial  \rho_{\beta}\vec{v}_{\beta} \over \partial t} + \nabla\cdot\left(\rho_{\beta} \vec{v}_{\beta}\vec{v}_{\beta}\right) 
+ \nabla P_{\beta}  +  \rho\nabla\Phi &=& 0 \ , \\
\nabla \cdot \nabla\Phi -4\pi G\left(\rho_{\alpha} + \rho_{\beta}\right)&=& 0 \ . 
\end{eqnarray}
Superseding, as in App. \ref{app:oc}, these equations with perturbation terms $A_{\alpha} = A_{\alpha 0} +
\delta A_\alpha$ and $A_{\beta} = A_{\beta 0} + \delta A_\beta$ in the $x$ direction and linearizing them yields
\begin{eqnarray}
{\partial\,{\delta\rho_\alpha} \over \partial t} + \rho_\alpha\nabla\cdot\delta v_\alpha &=& 0 \ ,  \\
{\partial\,\delta\rho_\beta \over \partial t} + \rho_\beta\nabla\cdot\delta v_\beta &=& 0 \ ,  \\
\rho_\alpha{\partial\,\delta v_\alpha \over \partial t} 
+ \left({\partial P_\alpha\over \partial \rho_\alpha}\right)_s \nabla \delta\rho_\alpha  
+ \rho_\alpha\nabla\delta\Phi&=& 0 \ , \\
\rho_\beta{\partial\,\delta v_\beta \over \partial t} 
+ \left({\partial P_\beta\over \partial \rho_\beta}\right)_s \nabla \delta\rho_\beta  
+ \rho_\beta\nabla\delta\Phi&=& 0 \ , \\
\nabla \cdot \nabla\delta\Phi - 4\pi G(\delta\rho_\alpha + \delta\rho_\beta)&=& 0 \ , \label{equ:lintp}
\end{eqnarray}
which transform into a linear equation system in Fourier space,
\begin{eqnarray}
-i\omega\,\hat{\rho}_\alpha + i k\,\rho_{\alpha 0} \hat{v}_\alpha &=& 0 \ , \\
-i\omega\,\hat{\rho}_\beta + i k\,\rho_{\beta 0} \hat{v}_\beta &=& 0 \ , \\
-i\omega\, \rho_{\alpha 0} \hat{v}_\alpha 
+  i k\,\left({\partial P_\alpha\over \partial \rho_\alpha}\right)_s \hat{\rho}_\alpha
+ i k \,\rho_{\alpha 0} \hat{\Phi} &=& 0 \ , \\
-i\omega\, \rho_{\beta 0} \hat{v}_\beta +  i k\,\left({\partial P_\beta\over \partial \rho_\beta}\right)_s \hat{\rho}_\beta
+ i k \,\rho_{\beta 0} \hat{\Phi} &=& 0 \ , \\
-k^2\hat{\Phi} - 4\pi G (\hat{\rho}_\alpha + \hat{\rho}_\beta) &=& 0 \ .
\end{eqnarray}
This can be written in the matrix form $\matr{A}\cdot\vec{x} = \vec{0}$, defining $c^2_\alpha = ({\partial P_\alpha /
        \partial \rho_\alpha})_s$ and $c^2_\beta = ({\partial P_\beta / \partial \rho_\beta})_s$ as follows:
\begin{equation}
\begin{bmatrix}
\omega & 0 & k\rho_{\alpha 0} & 0 & 0\\
0 & \omega & 0 & k\rho_{\beta 0} & 0\\
k\,c^2_\alpha & 0 & \omega\, \rho_{\alpha 0} & 0 & k \,\rho_{\alpha 0} \\
0 & k\,c^2_\beta & 0 & \omega\, \rho_{\beta 0} & k \,\rho_{\beta 0} \\
- 4\pi G& - 4\pi G & 0 & 0 & -k^2
\end{bmatrix} \cdot 
\begin{bmatrix}
\hat{\rho}_\alpha \\ \hat{\rho}_\beta \\ \hat{v}_\alpha \\ \hat{v}_\beta\\ \hat{\Phi}
\end{bmatrix}
= \begin{bmatrix}
0 \\ 0 \\ 0 \\ 0 \\ 0
\end{bmatrix} \ .
\end{equation}
In order to simplify, we set $\Gamma_\alpha = 4\pi G \rho_{\alpha 0}$ and $\Gamma_\beta = 4\pi G \rho_{\beta 0}$ and
find the following determinant:
\begin{multline}
\det(\matr{A}) = k^2 \rho_{\alpha 0} \rho_{\beta 0} \left[\omega^4  
+ \left(\Gamma_\alpha + \Gamma_\beta - k^2(c^2_\alpha + c^2_\beta)\right)\omega^2 \right.\\
\left.+ k^2\left(-\Gamma_\beta c^2_\alpha - \Gamma_\alpha c^2_\beta + k^2 c^2_\alpha c^2_\beta\right)\right] \ .
\end{multline}
Again, to have a non-trivial solution, its determinant must be zero, which, in the case of $k\neq 0$, is
\begin{equation}
\omega^4  + \left(\Gamma_\alpha + \Gamma_\beta - k^2(c^2_\alpha\! +\! c^2_\beta)\right)\omega^2 
- k^2\left(\Gamma_\beta c^2_\alpha + \Gamma_\alpha c^2_\beta - k^2 c^2_\alpha c^2_\beta\right) = 0 , \label{equ:gaka}
\end{equation}
with the following solution for $\omega^2$:
\begin{multline}
(\omega^2)_{1,2} = -{1\over 2}\left({\Gamma_\alpha+\Gamma_\beta-k^2(c^2_\alpha+c^2_\beta)}\right) \\ 
\pm{\sqrt{{1 \over 4}\left(\Gamma_\alpha+\Gamma_\beta- k^2(c^2_\alpha + c^2_\beta )\right)^2
+ k^2 \left(\Gamma_\beta c^2_\alpha +\Gamma_\alpha c^2_\beta - k^2 c^2_\alpha c^2_\beta\right)}} \ . 
\end{multline}
Setting $\omega^2 = 0$ in Equ.\ (\ref{equ:gaka}) yields
\begin{equation}
k^2\left(-\Gamma_\beta c^2_\alpha - \Gamma_\alpha c^2_\beta + k^2 c^2_\alpha c^2_\beta\right) = 0 \ ,
\end{equation}
with the following solution:
\begin{equation}
k^2_\name{GZ} \equiv {\Gamma_\alpha\over c^2_\alpha} +  {\Gamma_\beta\over c^2_\beta} \ ,
\end{equation}
a fluid is unstable for $\omega^2 < 0$ or $\omega^2\in \Im$, which is the case for $k^2 < k^2_\name{GZ}$.

\subsubsection{Phase transition}

In the case of a phase transition, one of the pressure derivatives is equal to zero. Setting $c_\alpha = 0$ in Equ.\
(\ref{equ:gaka}) we get
\begin{eqnarray}
\omega^4  
+ \left[\Gamma_\alpha + \Gamma_\beta -k^2 c^2_\beta\right]\omega^2 
-\Gamma_\alpha k^2 c_\beta^2 = 0\label{equ:ompt}  \ ,
\end{eqnarray}
and its solutions is written as
\begin{equation}
\omega^2 = -{1\over 2}\left({\Gamma_\alpha + \Gamma_\beta- k^2 c_\beta^2 }\right) \pm{\sqrt{{1\over 4}\left(\Gamma_\alpha 
	+ \Gamma_\beta-k^2 c_\beta^2\right)^2 + \Gamma_\alpha k^2  c_\beta^2}} \ . \label{equ:ompt2}
\end{equation}
Setting $\omega^2 = 0$ in  Equ.\ (\ref{equ:ompt}), only the trivial $k=0$ is a solution. Since
\begin{equation}
\left(\Gamma_\alpha+\Gamma_\beta-k^2 c_\beta^2\right)^2 < \left(\Gamma_\alpha+\Gamma_\beta-k^2 c_\beta^2\right)^2 
+ 4\,\Gamma_\alpha  k^2 c_\beta^2 \ , 
\end{equation}
the upper sign solution of Equ.\ (\ref{equ:ompt2}) is always positive and the lower sign solution is always negative for any $k$.
Therefore one $\omega$-solution of Equ.\ (\ref{equ:ompt2}) is always negative and thus unstable, independent of the
strength of either $\Gamma_\alpha$ or $\Gamma_\beta$.

\section{Energy and radiation transfer during the contraction of a sphere towards an ellipsoid}
\label{app:ppc}

\begin{figure}[t] 
    \centering
    \resizebox{\hsize}{!}{\includegraphics{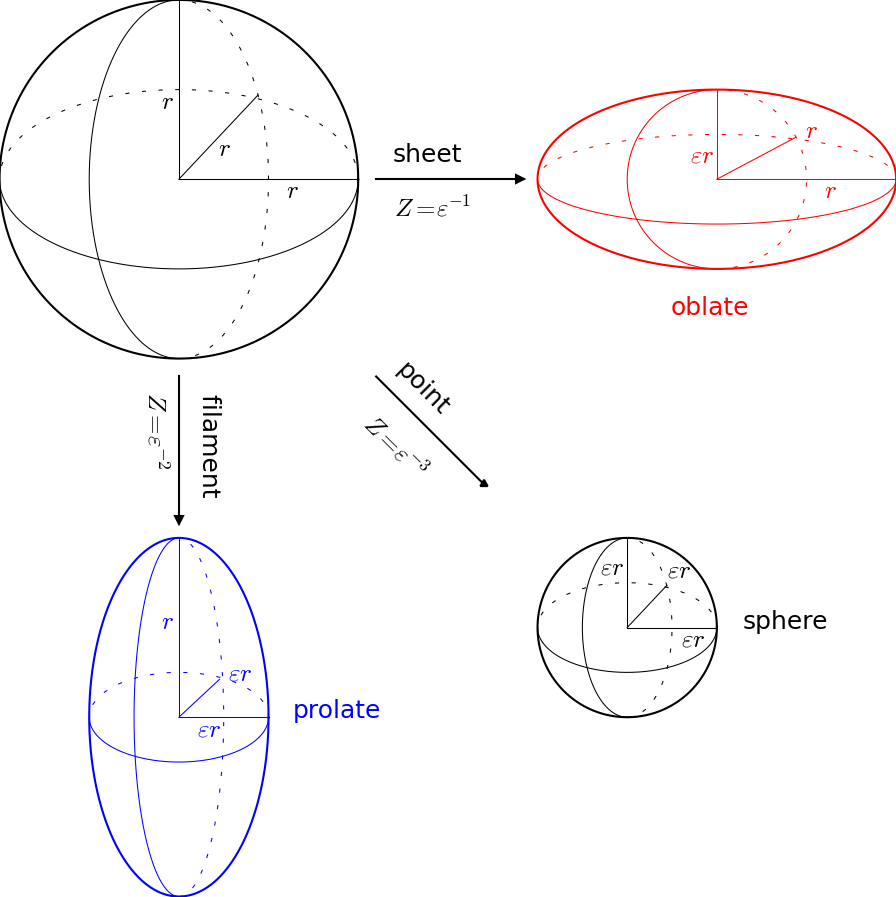}} 
    \caption{Collapsing geometries.}
    \label{fig:ell} 
\end{figure}

We consider a non-rotating sphere of radius $r$ initially in unstable equilibrium, which contracts at constant mass as
an ellipsoid with semi-principal axes $a$, $b,$ and $c$ (see Fig. \ref{fig:ell}).  In a sheet-like collapse, two
semi-axes remain the same ($a = b = r$) while one is decreasing (c = $\varepsilon r$), leading to an oblate spheroid. 
In a filament-like collapse, one semi-axis remains the same ($a = r$), while two are decreasing together ($b = c =
\varepsilon r$), leading to a prolate spheroid.  In a point-like collapse, all the three semi-axes decrease together ($a
= b= c=\varepsilon r$), remaining a sphere. During compression, density increases by a factor $Z = n/n_0$. Since the
ellipsoid volume is $V_\name{ell} = 4/3\,\pi a b c$, compression changes as: $\varepsilon_\name{oblate} = Z^{-1}$,
$\varepsilon_\name{prolate} = Z^{-1/2}$, and $\varepsilon_\name{sphere} = Z^{-1/3}$.

\subsection{Gravitational energy}
\label{app:ge}

The gravitational energy difference between the initial sphere and subsequent ellipsoids must be released as additional
thermal energy.  The gravitational energy of a revolution ellipsoid, with
$E_{G,0} = E_{G,\textrm{sphere}}(r) = -(3/5) GM^2/r$ \citep{landau_classical_1975}, is written as
\begin{eqnarray}
    {E_\textrm{oblate}(Z)\over E_{G,0}} &=& {\arccos\left(Z^{-1}\right) \over \sqrt{1-Z^{-2}}} 
    = {\pi \over 2} - Z^{-1} +O(Z^{-2}),\\
    {E_\textrm{prolate}(Z)\over E_{G,0}} &=& {\name{arcosh}\left(\sqrt{Z}\right) \over \sqrt{1-Z^{-1}}}
     = \log(2\sqrt{Z})+O\left( \log(Z) \over Z \right),\\
    {E_\textrm{sphere}(Z)\over E_{G,0}} &=&  Z^{1/3} .
\end{eqnarray}
Sheet-like contraction leads to infinite densities with \textit{finite} temperature increase, which is much more
favourable for reaching condensation conditions that filament-like or point-like contractions.

We show now that the maximum relative temperature increase of an initial perfect gas sphere initially in equilibrium is
bounded.  State 0 is the initial (unstable) equilibrium sphere case, and state 1 is any later, denser case that is not
necessarily in equilibrium.  Since in equilibrium, the initial state respects the virial condition,
\begin{eqnarray}
E_{G,0} + 2\, E_\name{kin,0} &=& 0 \ ,
\end{eqnarray}
where $E_\name{kin,0}$ is the kinetic energy.  Since at rest, the initial sphere kinetic energy consists only of
microscopic motion, and is proportional to the initial temperature $T_0$.

The initial and later total energies are,
\begin{eqnarray}
E_\name{tot,0} &=& E_{G,0} + E_\name{kin,0} \ , \\ 
E_\name{tot,1} &=& E_{G,1} + E_\name{kin,1} \ .
\end{eqnarray}
Taking into account possible radiative cooling, we suppose $E_\name{tot,1} \leq E_\name{tot,0}$, which leads to, using
the initial virial condition,
\begin{equation}
{T_1 \over T_0} \leq {E_\name{kin,1} \over E_\name{kin,0}} \leq 2\,{E_\name{G,1} \over E_\name{G,0}} - 1 \ .
\end{equation}
The first inequality takes into account that state 1 is not necessarily in equilibrium; some kinetic energy may be
attributed to macroscopic motion.

Thus, using the above potential energy ratios, in the case of an oblate spheroid contraction,
\begin{equation}
{T_1 \over T_0} \leq \pi -1 -2Z^{-1} + O(Z^{-2})\ ,
\end{equation}
that is, the final temperature cannot exceed $\pi-1\approx 2.1$ times the initial temperature.  In the case of a
prolate spheroid contraction, temperature is logarithmically bounded as $Z$ increases,
\begin{equation}
{T_1 \over T_0} \leq \log(4Z)-1  + O\left(\log(Z) \over Z\right)\ ,
\end{equation}
while in a spherical contraction, temperature is bounded by the cubic root of compression, 
\begin{equation}
{T_1 \over T_0} \leq 2Z^{1/3}-1  \ .
\end{equation}

\subsection{Radiative cooling}
\label{app:c}

Energy lost by radiation lowers temperature, but if opacity increases during contraction at some point the radiative
cooling rate drops below the heating rate as a result of gravitational energy conversion, thereby slowing down the
collapse.  Here we show with simple arguments how opacity changes when continuously contracting an initial sphere
towards denser, smaller spheres, or towards denser revolution of oblate or prolate kinds of ellipsoids, keeping the
longest axes constant and assuming  uniform densities at any stage and constant absorption cross sections.

\subsubsection{Optical depth}

The optical depth $\tau$ in the cumulated absorption over a photon path $\ell$: $\tau \equiv \int_0^\ell \sigma \,n\,
\dd \ell \ $, where $\sigma$ is the absorption cross section of individual atoms with number density $n$. The central
optical depth, calculated from the centre to the ellipsoid edge along some direction, is a first order estimator of the
average optical depth.  We compare the optical depth $\tau_0 = r\sigma\,n_0$ for the initial sphere with the later
spheres. For revolution ellipsoids, where $a$ and $c$ are the semi-long and short axes, respectively, the distance from
the centre to some point on the edge is $\ell(\theta) = ac / \sqrt{a^2\sin^2\theta + c^2\cos^2\theta}$ for oblate
spheroids and $\ell(\theta) = ac/\sqrt{a^2\cos^2\theta + c^2\sin^2\theta}$ for prolate spheroids. The angle $\theta$
vanishes at the spheroid equator.  Since the ellipticity $\varepsilon=c/a$ varies as $Z^{-1}$, $Z^{-1/2}$, and
$Z^{-1/3}$ in the oblate, prolate, and spherical cases, respectively, the optical depth ratios as functions of
compression $Z$ and $\theta$ are found to be 
\begin{eqnarray}
    {\tau_\name{oblate} \over \tau_0} &=& {1 \over \sqrt{ \sin^2\theta + Z^{-2}\cos^2\theta }} \ , \\
    {\tau_\name{prolate}\over \tau_0} &=& {Z^{1/2} \over \sqrt{\cos^2\theta + Z^{-1} \sin^2\theta}} \ , \\
    {\tau_\name{sphere} \over \tau_0} &=& Z^{2/3} \ .
\end{eqnarray}
Thus, in the oblate case the central optical depth ratio does not change along the poles and at high compression remains
barely increased over most directions.  In the prolate case it increases  least along the equator, but is proportional
to the square root of compression.  In the spherical case it increases  most rapidly as a power $2/3$ of compression. 
Thus sheet-like compression provides the least optical depth increase and spherical compression compression the most.

\subsubsection{Global absorption}

\begin{figure}[t] 
    \resizebox{\hsize}{!}{\includegraphics{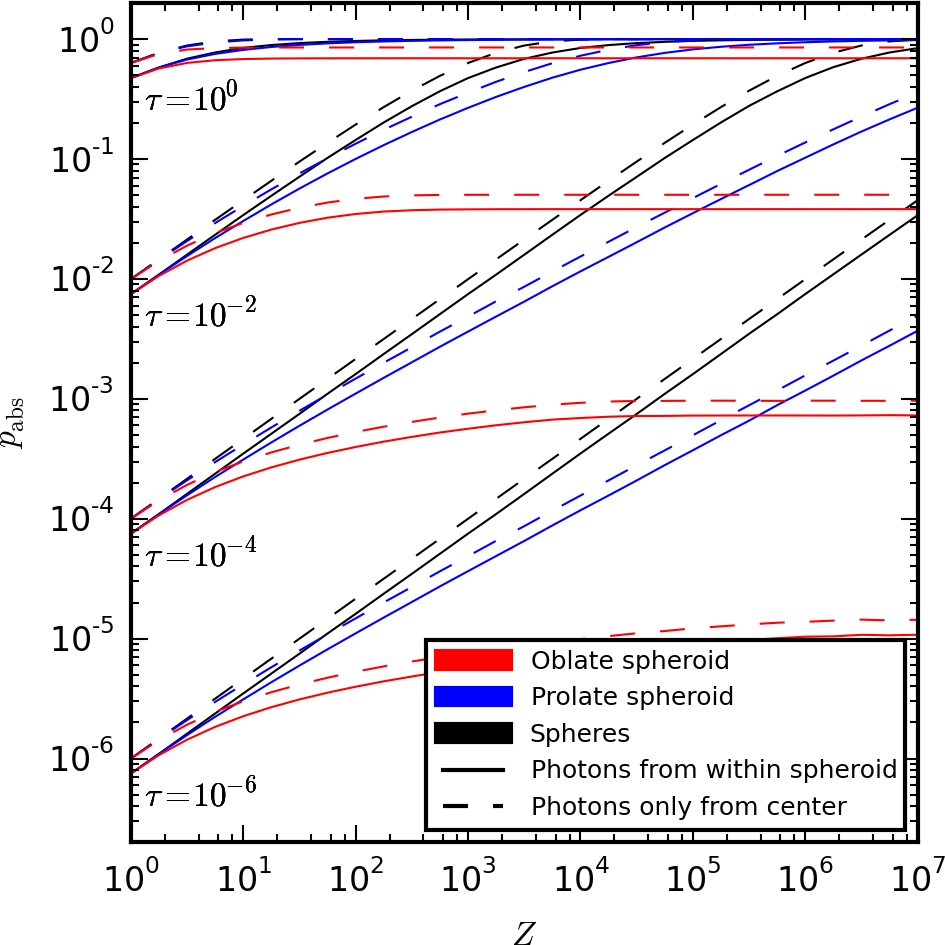}} 
    \caption{Absorption probability in contracting spheroids as function of compression $Z>1$ calculated by Monte Carlo
        simulation. At $Z=1$ all cases are spherical. The photons start either at the centre only or anywhere inside the
        ellipsoid in random directions.  Different cases are represented where the initial sphere optical depth $\tau_0$ is indicated.}
    \label{fig:abs} 
\end{figure}

One can refine the previous estimate for cooling by calculating, for any point inside an ellipsoid, the probability for
a photon to be absorbed. For a given optical depth $\tau$ the absorption probability is $p = 1 - \exp(-\tau)$.  The
global probability of absorption must be calculated for all solid angles for all points.  These 4- or 5-dimensional
integrals for bi- or tri-axial ellipsoids does not seem to be solvable analytically, and straightforward numerical
quadratures would be expensive.  Thus we resort to a Monte Carlo draw to estimate these quantities.  We pick randomly
and uniformly a number of points inside the ellipsoid and a random, uniform directional unit vector $\vec n$, and find
the two distances $\ell_1$, $\ell_2$, to the edge of the ellipsoid, allowing us to calculate two optical depths
$\tau_1$, $\tau_2$, and the corresponding absorption probabilities $p_1$, $p_2$ for each point.  Knowing the starting
position $\vec x$ inside the ellipsoid $(a,b,c)$ and the direction vector $\vec n$, we find the two signed distances to
the ellipsoid edge by solving the quadratic equation $(x+\ell n_x)/a^2+(y+\ell n_y)^2/b^2+(z+\ell n_z)^2/c^2=1$ for
$\ell$. Explicitly,  noting $\alpha= a^{-2}$, $\beta= b^{-2}$, $\gamma=c^{-2}$, for each point $\vec x = [x,y,z]$ the
procedure becomes
\begin{eqnarray}
  A &=& \alpha n_x^2 + \beta n_y^2 + \gamma n_z^2 \ , \\
  B &=& \alpha x n_x + \beta y n_y + \gamma z n_z \ , \\
  C &=& \alpha x^2 + \beta y^2 + \gamma z^2 - 1 \ , \\
  D &=& \sqrt{B^2-AC}, \\
  \ell_1 &=& -(D+B)/A ,\quad \ell_2 ~=~ (D-B)/A \ . 
\end{eqnarray}
For each set of $\ell_i$, average absorption probabilities can be found for a range of $\sigma$s.  The two absorption
probabilities $p_i=1-\exp(\sigma n|\ell_i|)$, $i=1,2$, provide two distinct probabilities for each point.  Each set of
$p_i$s should converge towards a similar average value.  The difference allows us to check the error obtained with a
finite number of points.  Between $2\cdot 10^4$ (sphere case) and $3\cdot 10^7$ points (oblate spheroid case) were drawn
such that the $\log_{10} p_i$ between the two sets differ by at most 0.01.  The result is shown in Fig.\ \ref{fig:abs}.
The error bars are comparable or smaller than the thickness of the curve. 
\medskip

The sphere and prolate spheroid cases quickly become optically thick, increasing as $Z^{2/3}$ and $Z^{1/2}$,
respectively, in the optically thin regime.  In contrast, the absorption of a contracting optically thin oblate spheroid
increases logarithmically until it reaches $Z \tau_0 {\sim} 1$ beyond which it remains approximately constant; in other
words if the initial state is optically thin, it remains so even after infinite compression.  The emission signature of
a collapsing sheet should therefore remain observationally barely noticeable, since both temperature and optical
thickness increase by very modest factors in comparison with the other geometries.

Fig.\ \ref{fig:maxlab} shows how an initial sphere at $T = 10\,\unit{K}$, $P = 10^{-12}\,\unit{Pa}$ would change its
temperature and pressure when contracting adiabatically, changing its gravitational energy into thermal energy.  Clearly
the sheet-like collapse is the most favourable geometry for reaching the H$_2$ phase transition regime. Including
radiative cooling, which is the fastest in sheet-like geometry, can only help in this regard.

\end{appendix}

\end{document}